\documentclass{JINST}
\usepackage{lineno}
\usepackage[percent]{overpic}
\usepackage{amsmath}
\usepackage{amssymb}
\newcommand{\DBD}{0$\nu$DBD}
\newcommand{\PO}{$^{210}$Po}

\newcommand{\TEO}{$\mathrm{TeO}_2$}
\newcommand{\TEHT}{$^{130}\mathrm{Te}$}

\newcommand{\THO}{$^{232}\mathrm{Th}$}

\newcommand{\Ohm}{\un\Omega}
\newcommand{\MO}{\un{M\Ohm}}
\newcommand{\GO}{\un{G\Ohm}}

\newcommand{\Cuore}{CUORE}

\providecommand*{\un}[1]{\ensuremath{\mathrm{\,#1}}}

\title{Signal and noise simulation of CUORE bolometric detectors}
\author{
M.~Carrettoni$^a$ and M.~Vignati$^b$\thanks{Corresponding author.}\\
\llap{$^a$}Dipartimento di Fisica, Universit\`a di Milano-Bicocca, Milano I-20126, Italy\\
\llap{$^b$}Dipartimento di Fisica, Sapienza Universit\`a di Roma and Sezione INFN di Roma, Roma I-00185, Italy\\
E-mail:\email{marco.vignati@roma1.infn.it}
}

\abstract{
Bolometric detectors are used in particle physics experiments to
search for rare processes, such as neutrinoless double beta decay and dark
matter interactions. By operating at cryogenic temperatures, they are able
to detect particle energies from a few keV up to several MeV, measuring
the temperature rise produced by the energy released. This work focusses on the
bolometers of the CUORE experiment, which are made of TeO$_2$ crystals. The response of these detectors is
nonlinear with energy and changes with the operating temperature. The
noise depends on the working conditions and significantly affects the
energy resolution and the detection performances at low energies.
We present a software tool to simulate  signal
and noise of CUORE-like bolometers, including effects generated by operating temperature drifts,
nonlinearities and pileups. The simulations  agree well
with data.
}

\begin{document}

\section{Introduction}
Bolometers are detectors in which the energy from particle interactions is converted
to heat and measured via their rise in temperature. They provide excellent
energy resolution, though their response is slow compared to electronic or photonic detectors.
These features make them a suitable choice for experiments searching for rare
processes, such as neutrinoless double beta decay (\DBD) and dark matter (DM)
interactions.

The \Cuore\ experiment will search for \DBD\ of
\TEHT~\cite{Ardito:2005ar, ACryo} using an array of 988
\TEO\ bolometers of 750\un{g} each. It may also be sensitive
to DM interactions~\cite{DiDomizio:2010ph}. 
Operated at a temperature
of about 10\un{mK}, these detectors exhibit an energy resolution
of a few keV over an energy range extending from a few keV up
to several MeV.  In this range the response function is found to be
nonlinear~\cite{Vignati:2010yf}.  The conversion from signal amplitude to
energy is complicated and the shape of the signal depends on the energy
itself.  Moreover, the amplitude of the signal depends on the temperature
of the detector, which is very difficult to keep stable with current cryostats within the  few ppm level, the level that would not perturb the energy resolution.  The noise of the detector is
dominated by thermal fluctuations induced by vibrations, and significantly
affects the energy resolution at low energies~\cite{Bellini:2010iw}.

In this paper we present a method to simulate signal and noise of
CUORE-like bolometers. The simulation should be able to reproduce all
the features of the data and can be used, for example, to estimate
detection efficiencies and to test analysis algorithms.  The shape of
the signal and the noise are estimated from the data.  The nonlinearities
of the signal are reproduced using a model of the thermal sensor of the
bolometer~\cite{Vignati:2010yf}.

\section{Description of the detector} 
\begin{figure}[htbp]
\begin{center}
\includegraphics[clip=true,width=0.6\textwidth]{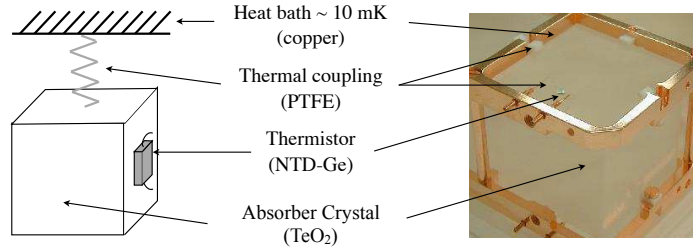}
\caption{Sketch of a \Cuore-like\ bolometer (left) and a photograph of a bolometer
(right). The \TEO\ crystal is held by Teflon supports, the thermistor
is glued to the crystal and its wires are attached to the copper frame. The
supports and the thermistor wires thermally couple the crystal to the
copper frame, which act as heat bath. } 
\label{fig:crystal_setup}
\end{center}
\end{figure}

A \Cuore\ bolometer is composed
of two main parts, a \TEO\ crystal and a neutron transmutation doped
Germanium (NTD-Ge) thermistor~\cite{wang,Itoh}. The crystal is cube-shaped
(5x5x5\un{cm^3}) and held by Teflon supports in copper frames.  The frames
are coupled to the mixing chamber of a dilution refrigerator, which
keeps the system at a temperature of $\sim 10\un{mK}$.  The thermistor
is glued to the crystal and acts as thermometer (Fig.~\ref{fig:crystal_setup}). %
A Joule heater is also glued to most crystals.  It is used to
inject controlled amounts of energy into the crystal, to emulate signals
produced by particles~\cite{stabilization,Arnaboldi:2003yp}.  When energy
is released in the crystal, the crystal temperature increases and changes the
thermistor's resistance according to the relationship~\cite{Mott:1969}:
\begin{equation}
R(T) = R_0\exp \left(T_0 / T \right)^\gamma
 \label{eq:thermistor}
\end{equation}
where $R_0$ and $T_0$ are parameters that depend on the dimensions and
on the material of the thermistor.  For \Cuore\ bolometers values are
about $1.1\un{\Ohm}$ and $3.4\un{K}$, respectively. At 10\un{mK}
the parameter $\gamma$ can be considered constant and equal to $1/2$
~\cite{efros,Itoh:1996}.

\begin{figure}[htbp]
\begin{center}
\includegraphics[clip=false,width=0.5\textwidth]{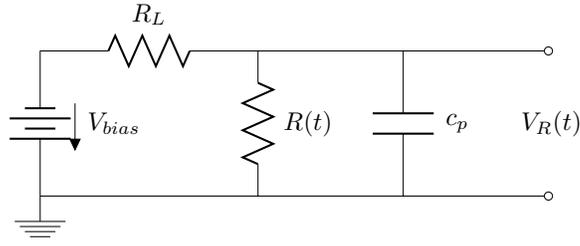}
\caption{Biasing circuit of the thermistor. A voltage generator $V_{bias}$
biases the thermistor resistance $R(t)$ in series with a load resistance
$R_L$. The bolometer signal is the voltage $V_R(t)$ across $R(t)$. The
wires used to extract $V_R(t)$ from the cryostat have a non-negligible capacitance $c_p$.
}
\label{fig:biasing}

\end{center}
\end{figure}

To read out the signal, the thermistor is biased in differential configuration
with a bipolar voltage generator $\pm V_{bias}$ connected to a pair of load resistors,
$R_L$'s, finally connected to the thermistor's terminals.  The resistance of the thermistor
varies in time with the temperature, $R(t)$, and the voltage across it,
$V_R(t)$, is the bolometer signal.  The value of $R_L$'s is chosen to
be much higher than $R(t)$ so that  $V_R(t)$ is proportional
to $R(t)$.  Since from Eq.~(\ref{eq:thermistor}) positive temperature
variations induce negative resistance variations, the polarity of $V_{bias}$ is
chosen to be negative in order to obtain positive signals.  The connecting
wires add in parallel to the thermistor a parasitic  capacitance $c_p$. A schematic
of the biasing circuit is shown in Fig.~\ref{fig:biasing}. In  the figure, the series of the two load resistors
is represented as a unique resistor $R_L$. The signal $V_R(t)$ is amplified, filtered
with a 6-pole active Bessel filter, and then digitized with an 18-bit
analog-to-digital converter (ADC). To fit the signal in the range of the ADC, which is [-10.5,10.5]\un{V}, 
a programmable offset voltage $V_h$ is added to $V_R(t)$.  The front-end electronics, which
provide the bias voltage, the load resistors,  the amplifier and the offset voltage, are placed
outside of the cryostat, at ambient temperature~\cite{AProgFE}.

At 10\un{mK} the value of $R(T)$ is of order $100\MO$, $R_L$ is chosen
as $54\GO$ ($27 + 27\GO$) and $V_{bias}$ as $\sim 5$\un{V}.  The $c_p$ value depends
on the length of the wires that carry the signal out of the cryostat,
typically it is of order 400\un{pF}. The amplifier gain, the Bessel
filter frequency bandwidth, the duration of the acquisition window and the
sampling frequency are set typically at 5000\un{V/V}, 12\un{Hz},
5.008\un{s} and 125\un{Hz}, respectively.

The data analyzed in this paper come from test bolometers operated by the
\Cuore\ collaboration at the Gran Sasso underground laboratory (LNGS) in
Italy~\cite{ioanprod}.  The bolometers were  exposed to a $^{232}$Th 
calibration source
which, together with an $\alpha$ line generated by \PO\
contamination in the crystal, allows the analysis of an energy range
up to 5407\un{keV}.  

\begin{figure}[htbp]
\begin{center}
\includegraphics[clip=true,width=0.7\textwidth]{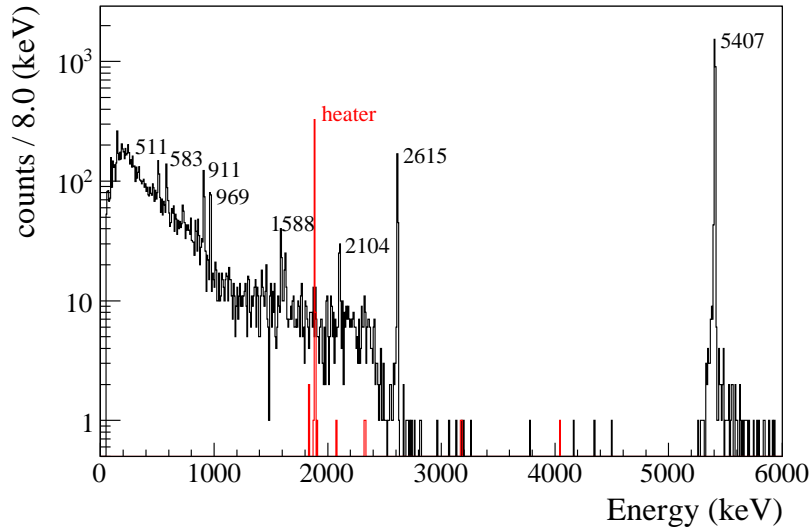}
\caption{Energy spectrum. All lines
are generated by the \THO\ calibration source except for the line at
5407\un{keV}, arising from $^{210}$Po contamination in the \TEO\ crystal.
Heater pulses were fired at an energy of $1885\un{keV}$.}
\label{fig:energy_spectrum}
\end{center}
\end{figure}

To simplify the description of this work we will focus our analysis on a single
bolometer.  The signal
rate on that bolometer was 133 mHz, that has to be combined to the rate
of heater pulses that were fired at an energy of 1885\un{keV} every 300
seconds (3.3 mHz).  The energy spectrum acquired in about $3\un{days}$
is shown in Fig.~\ref{fig:energy_spectrum}.
%

\section{Signal and noise features}\label{sec:snfeatures}

\begin{figure}[htbp]
\centering
\begin{minipage}{0.48\textwidth}
\includegraphics[clip=true,width=1\textwidth]{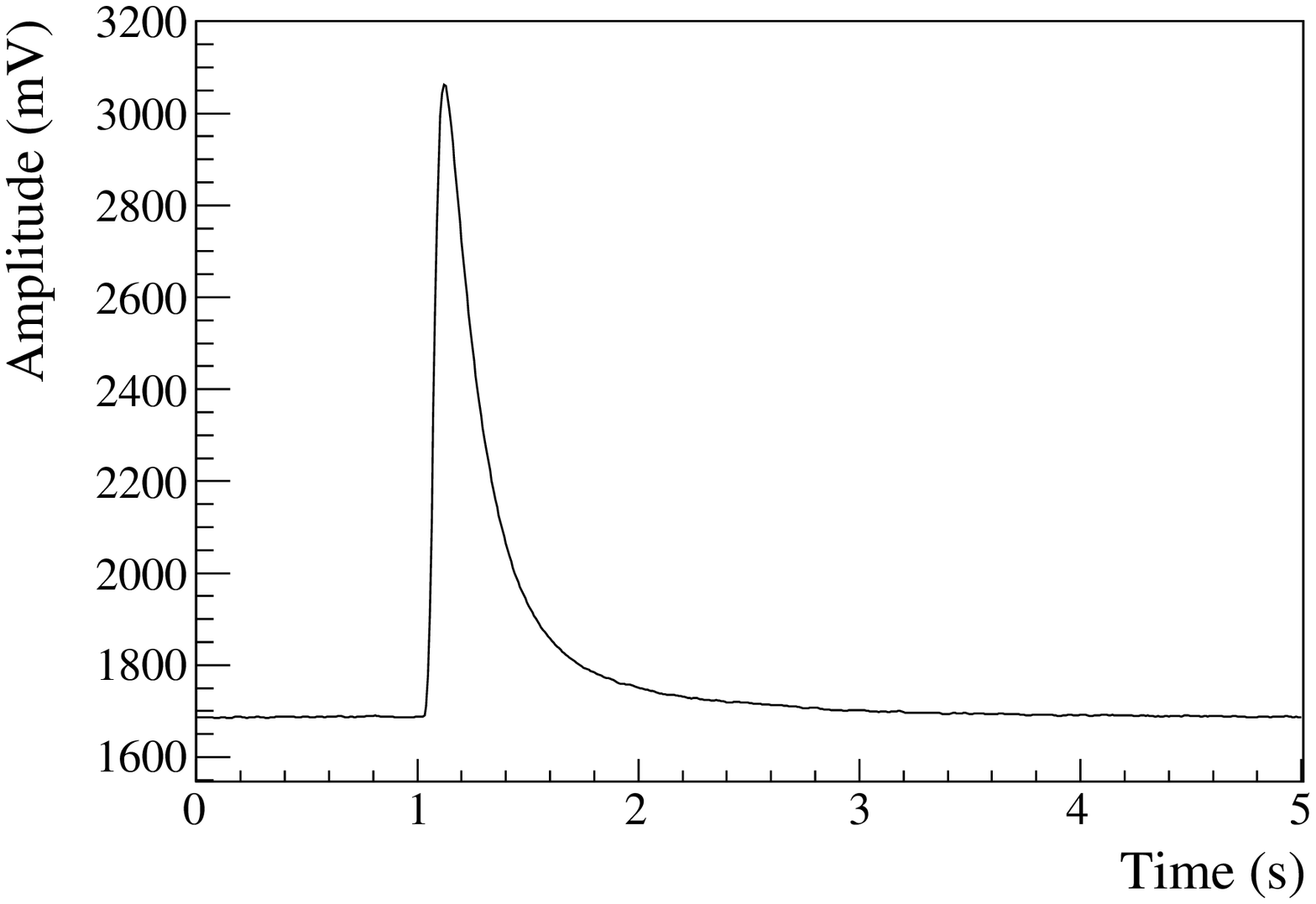}  
\end{minipage}
\hfill
\begin{minipage}{0.48\textwidth}
\includegraphics[clip=true,width=1\textwidth]{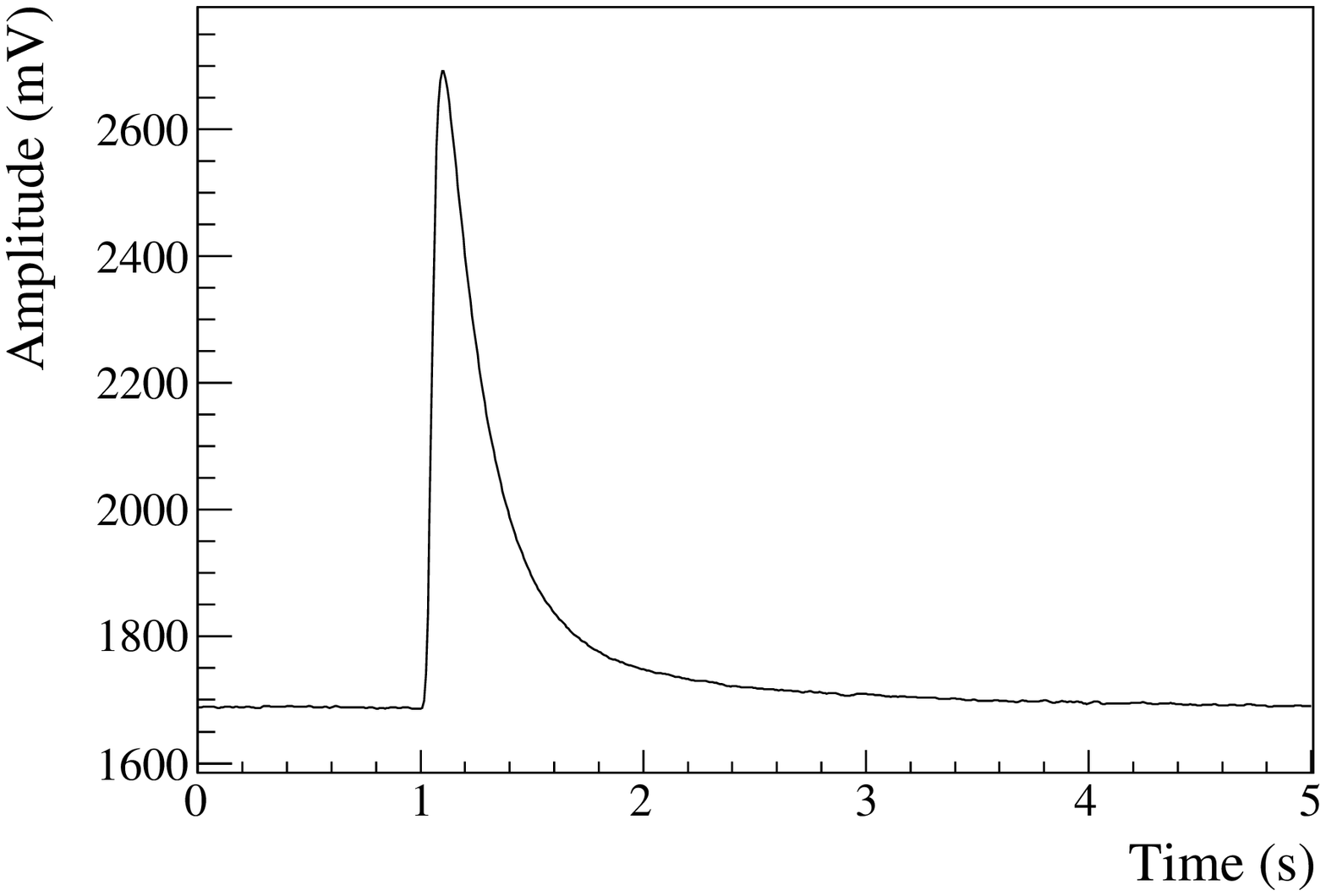}  
\end{minipage}
\caption{Pulse shapes of a 2615\un{keV} $\gamma$-ray
(left) and heater (right). The baseline is related to the temperature
of the thermistor before the particle interaction or the heater shot.
The amplitude carries information on the amount of energy released.}
\label{fig:pulses}
\end{figure}

Examples of signals generated by a 2615\un{keV} $\gamma$-ray
and by the heater, as acquired by the ADC, are shown in
Fig.~\ref{fig:pulses}. The baseline voltage of the pulses
is related to the thermistor temperature in static conditions, and
the amplitude is related to the energy released.  The shape of heater
pulses is found to be slightly different from that of particle pulses:
the rise time of the particle (heater) pulse in the figure,
computed as the time difference between the 10\% and the 90\% of the leading edge, 
is 55 (54) \un{ms} while the decay time, computed as the difference
between
the 90\% and 30\% of the trailing edge, is 220 (255) \un{ms}.

As already observed in Ref.~\cite{Vignati:2010yf} several nonlinearities are
present:
\begin{enumerate}
\item {The rise and the decay times of a pulse depend on the energy (Fig.~\ref{fig:shape_std}). 
}
\begin{figure}[btp]
\centering
\begin{minipage}{0.48\textwidth}
    \begin{overpic}[clip=true,width=1\textwidth]{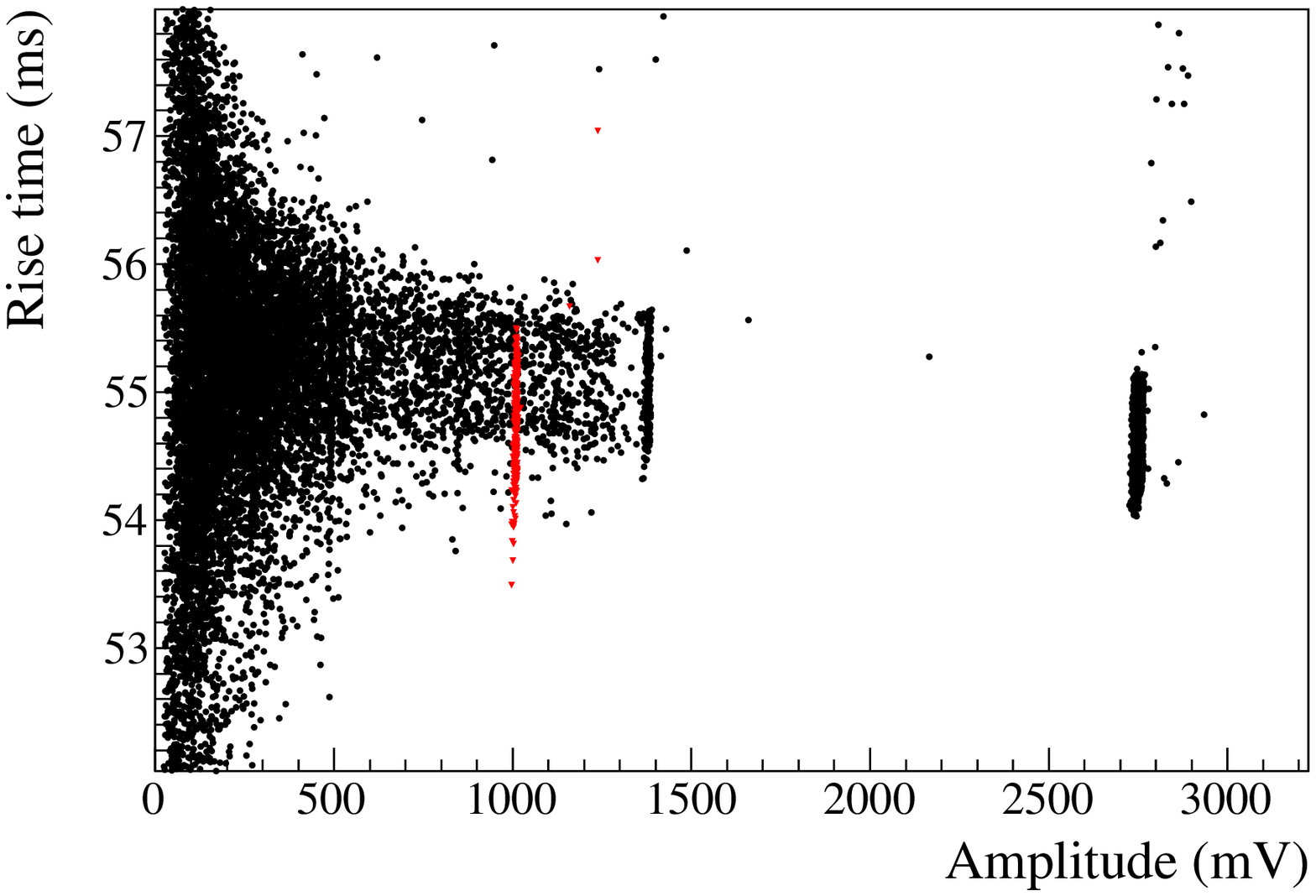}  
    \put(34,20){\footnotesize heater}
    \end{overpic}
\end{minipage}\hfill
\begin{minipage}{0.48\textwidth}
   \begin{overpic}[clip=true,width=1\textwidth]{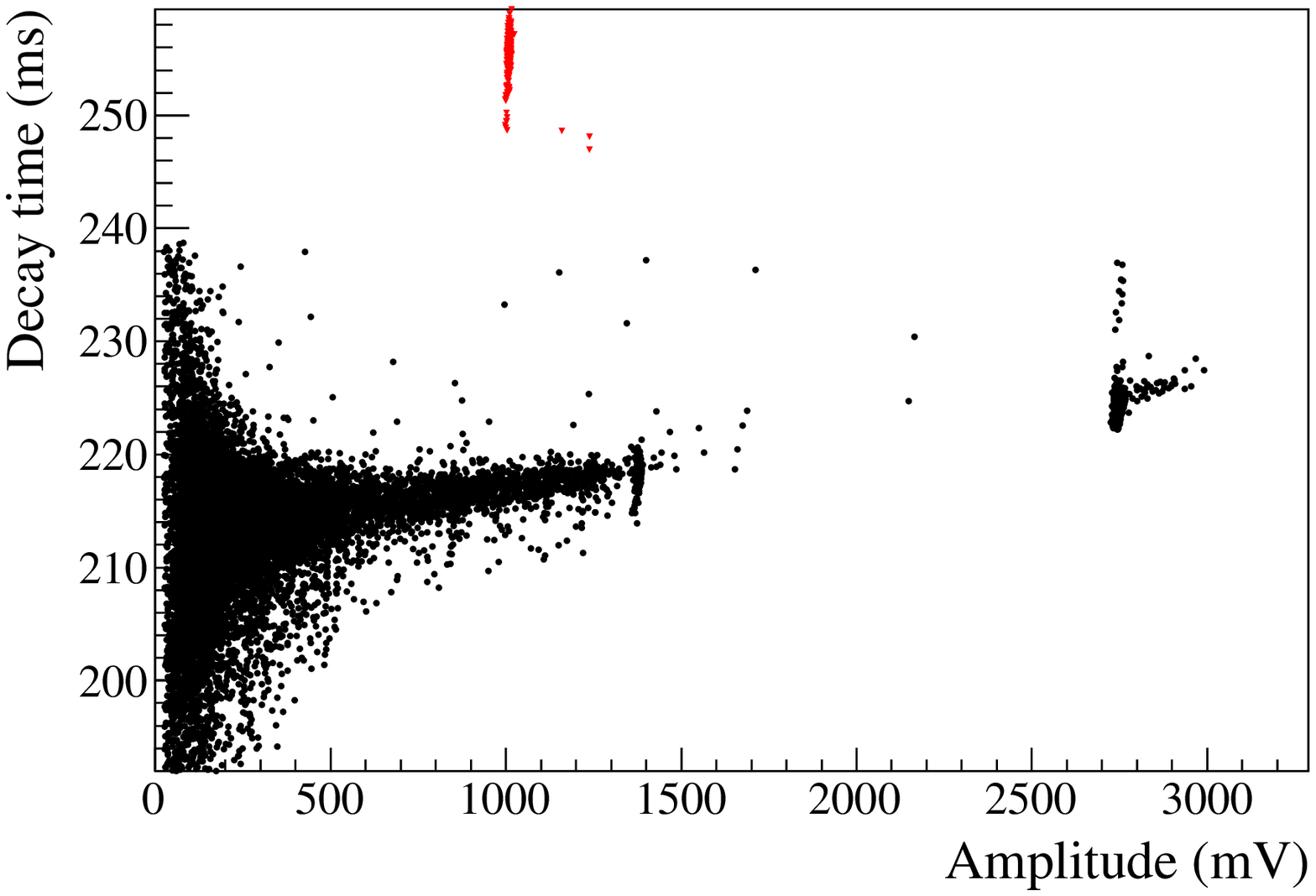}
    \put(42,60){\footnotesize heater}
   \end{overpic}  
\end{minipage}
\caption{Pulse shape parameters versus energy. 
The correlation with energy is negative for the rise time (left) and
positive for the decay time (right). Heater pulses are marked in red.} \label{fig:shape_std}
\end{figure}
\item { The amplitude of the pulse depends on the base temperature,
which varies during the data acquisition (Fig.~\ref{fig:stab_std}).
\begin{figure}[htbp]
\begin{center}
\includegraphics[clip=true,width=0.5\textwidth]{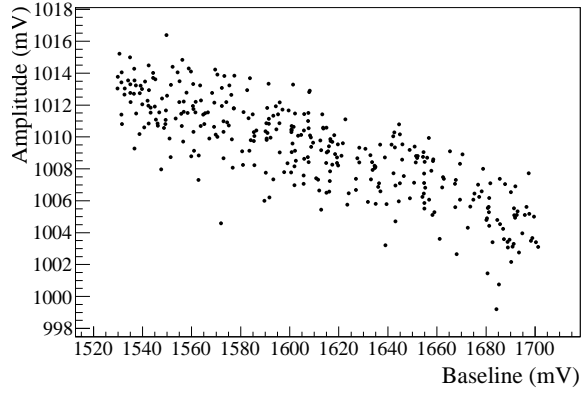}
\caption{Amplitude of heater pulses versus baseline. A change in
the bolometer temperature also changes its response, degrading the
energy resolution.} \label{fig:stab_std}
\end{center}
\end{figure}
}
\item { The amplitude dependence on energy is not linear. The deviation
from linearity of the data is estimated by comparing the result of a linear
calibration function,
\begin{equation}
\rm{Energy} = \rm{constant}\cdot\rm{Amplitude}\;,
\end{equation}
to the true energy of the source peaks.  The residuals evaluated
on the peaks generated by the $^{232}$Th source (see Fig.~\ref{fig:energy_spectrum}) are shown in Fig.~\ref{fig:res_std}.
The 5407\un{keV} line is not considered here because $\alpha$
particles have a quenching factor different from $\gamma$ and $\beta$
particles~\cite{Bellini:2010iw}.
\begin{figure}[tbp]
\begin{center}
\includegraphics[width=0.5\textwidth]{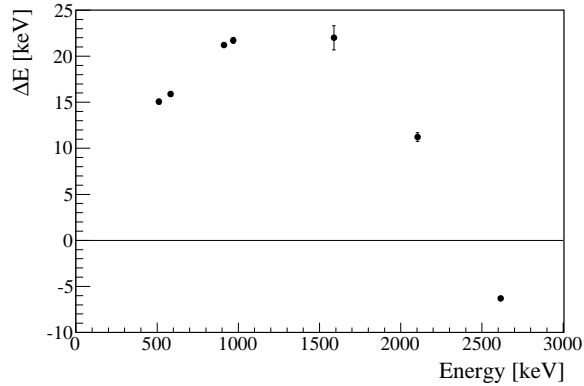}
\caption{Residuals obtained using a linear calibration function
for the well-identified peaks in the $^{232}$Th source spectrum. 
Considering that  the energy resolution is $\sim5\un{keV\,FWHM}$,
the difference $\Delta E$  between the estimated energies
and the true peak energies is not compatible with zero. The error bars refer to the
uncertainty on the estimated peak position that depends on the FWHM
and on the number of events $N$ in the peak as ${\rm FWHM} / (2.35\sqrt{N})$.}
\label{fig:res_std}
\end{center}
\end{figure}
}
\end{enumerate}

\begin{figure}[htbp]
\centering
\begin{minipage}{0.48\textwidth}
\includegraphics[clip=true,width=1\textwidth]{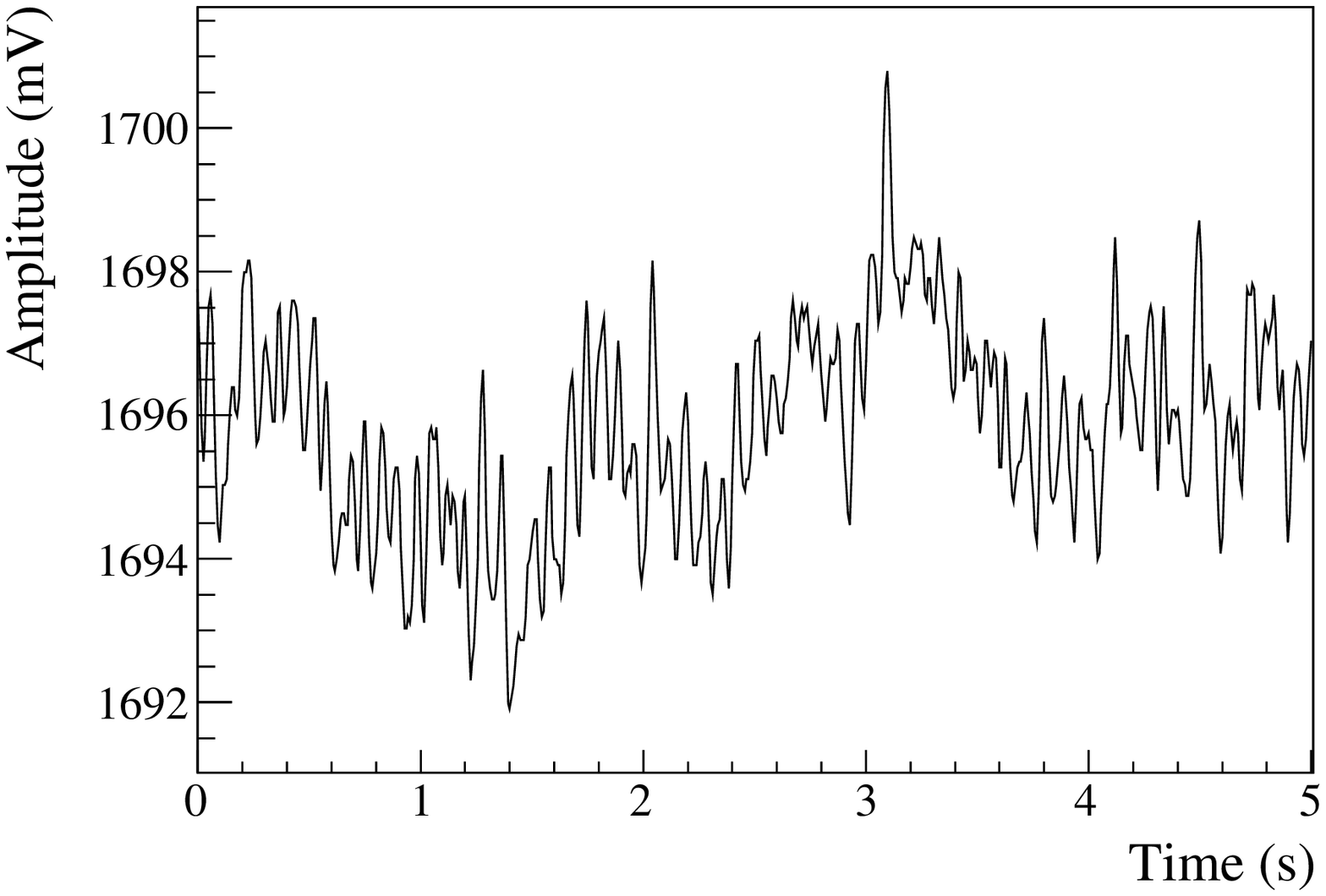}  
\end{minipage}\hfill
\begin{minipage}{0.48\textwidth}
\includegraphics[clip=true,width=1\textwidth]{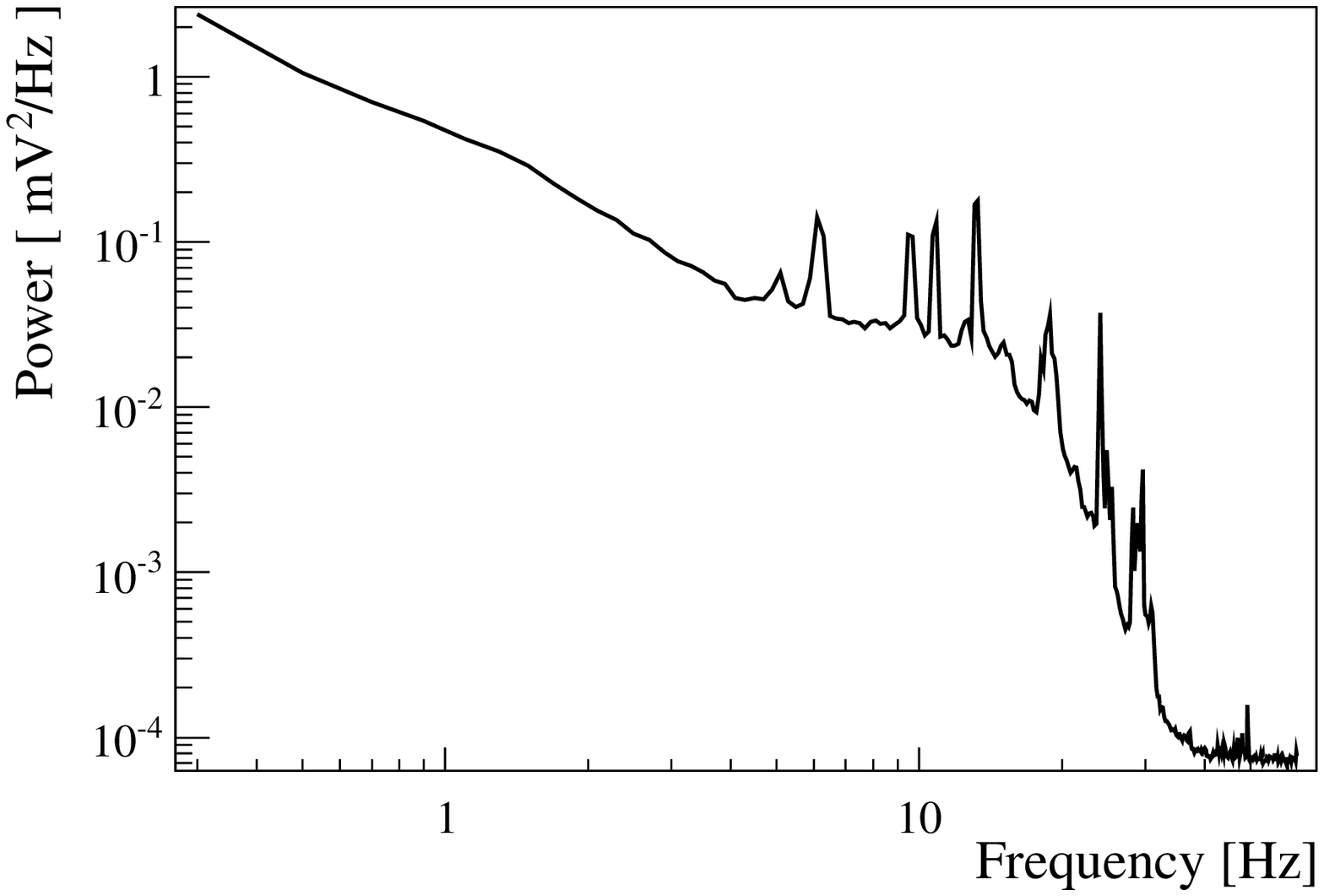}  
\end{minipage}
\caption{Noise of a bolometer at the output of the acquisition chain. A sample waveform (left) and
the power spectrum estimated from a large number of waveforms (right).}
\label{fig:noise_std}
\end{figure}

The noise of the bolometer in the signal frequency region (0-10\un{Hz})
is a result of vibrations inside the cryostat, and depends on the mechanical
setup of the experimental apparatus. The noise power spectrum was estimated
as:
\begin{equation}
N(\omega_k)\,=\;<|n(\omega_k)|^2>
\label{eq:nps}
\end{equation}
where $n(\omega_k)$ is the k-th component of the discrete Fourier
transform (DFT) of an acquired waveform not containing signals, and
$<>$ denotes the average over a large number of waveforms.
A typical noise waveform and the estimated power spectrum
are shown in Fig.~\ref{fig:noise_std}. 
The peaks in the power spectrum are due to the crystal friction
against its frame, and the residual common mode contribution 
from the vibration of the connecting wires (readout by the differential preamplifier).
Which of the two sources is dominant may depend on the set-up.
The white noise contribution at high frequency is due to the ADC digitization.

In the next sections we will describe the procedure we
developed to simulate the signal shape, the nonlinearities and the noise.
%

\section{Signal model} 

To simulate the data we developed a model of the bolometer signal that
reproduces the pulse shape and the nonlinearities. Starting from the
energy release in the crystal, the model is divided into three main stages:
the thermal response of the bolometer, the response of the
thermistor and of its biasing circuit, and the simulation of the electronics. 
All parameters are detector variables, except for the thermal model parameters, 
which we determine from fits to the data.

\subsection{Thermal model}\label{sec:thermalmodel}

\begin{figure}[htbp]
\begin{center}
\includegraphics[width=0.9\textwidth,clip=false]{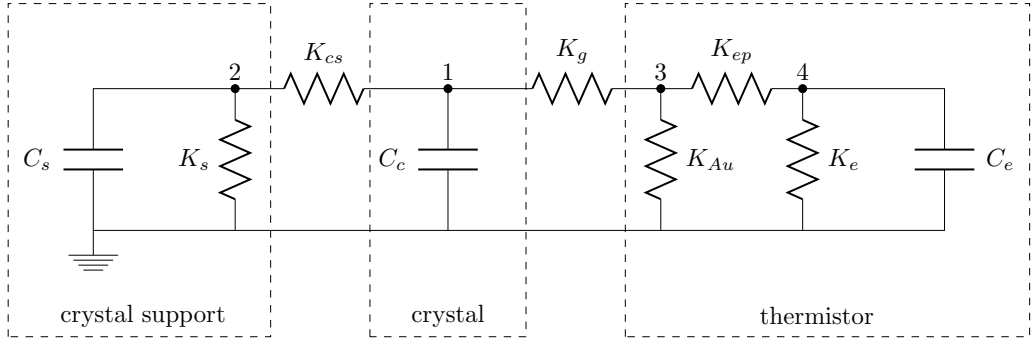}
\caption{Thermal circuit of a CUORE-like bolometer. Each of the elements is identified in the text.}
\label{fig:bolometer_circuit}
\end{center}
\end{figure}

A bolometer is a thermal system composed of a crystal, crystal supports,
thermistor, and their coupling elements~\cite{pessinamod} (Fig.~\ref{fig:bolometer_circuit}).
The crystal capacitance $C_c$ is connected to the thermistor through the glue spots
with conductance $K_g$ and to the supports through a contact conductance
$K_{cs}$.  The thermistor can be represented as a two-stage system
composed of a lattice and an electron gas, each with its 
capacitance and conductance, and inter-connected by a conductance $K_{ep}$. The
lattice capacitance, not shown in the figure, is negligible and it
discharges through the gold wires connected to the main heat bath
$K_{Au}$.  The electron gas capacitance and conductance are labeled as
$C_{e}$ and $K_{e}$, respectively.
The left side of the circuit shows the crystal supports 
with their capacitance $C_s$ and 
their conductance to the main bath
$K_s$. The main bath acts as the reference ground.

We are interested in the expression of the temperature variation of the thermistor electron gas
(node 4 in Fig.~\ref{fig:bolometer_circuit}) as a function of the time $\Delta T(t)$ after that an amount of energy $E$
is released in the crystal. The energy release 
is effectively instantaneous, because the phonons produced by a particle
 thermalize in a time much shorter  than the rise
time of acquired pulses~\cite{probst:1995}.  Under these conditions
the analytical expression of $\Delta T(t)$ is found to be:
\begin{equation}
\Delta T(t) = A_T \left(-e^{-\frac{t}{\tau_r}} + \alpha e^{-\frac{t}{\tau_{d1}}} + (1-\alpha) e^{-\frac{t}{\tau_{d2}}} \right)\;,
\label{eq:thermal_pulse}
\end{equation}
where all the parameters are complicated functions of the elements of the thermal circuit.  
The formula indicates that the thermistor temperature increases with one rise time constant, $\tau_r$, and 
  decreases with  two decay constants, $\tau_{d1}$ and $\tau_{d2}$.  
The parameter $\alpha$ weighs the two exponential decays and satisfies the condition $0\leq\alpha\leq1$.
The amplitude of the thermal pulse, $A_T$,  is also a function of the thermal elements
and is directly proportional to $E$.

In principle the parameters in Eq.~\ref{eq:thermal_pulse} could
be known if one were able to measure the underlying thermal
elements. 
Previous investigators have measured the thermal parameters~\cite{pessinamod},
and obtained heat capacitances and heat conductances of order
$10^{-a}\un{J/K}$ and $10^{-b}\un{W/K}$ with $a$ in the range 9 to 10 and $b$ in the range
9 to 11. Such measurements are of little utility for us for several reasons.
First, the capacitances and conductances depend on
the temperature and should be measured in the bolometer working
temperatures, that depend on the setup. Second, several thermal elements
vary from bolometer to bolometer.  For example, the contact heat conductance
between the crystal and its supports ($K_{cs}$) changes with the
detector configuration, and the heat conductance of the glue ($K_g$)
varies because of the weak reproducibility of the glue deposition. 
Third, we should know the parameters with a 
precision of the order of the present energy
resolution ($0.1\%$), which is not feasible using ordinary measurements techniques. 
We choose instead to estimate the parameters in Eq.~\ref{eq:thermal_pulse} 
from fits to the signal waveforms (see Sec.~\ref{sec:pulsefit}).

The solution of the circuit as expressed in Eq.~\ref{eq:thermal_pulse} does not
include the  dependence of the thermal elements on the
temperature. This potential
source of nonlinearity, however, is not visible in the
data~\cite{Vignati:2010yf}.  We also neglected the electrothermal
feedback effect:  when temperature change induces resistance change
in a biased thermistor, the resulting excursion in Joule heating
induces further temperature change. It can be shown that the electrothermal feedback
to the first order acts as a correction to the values of $K_{e}$ and
$C_{e}$, and therefore does not affect the configuration of the thermal
circuit and the form of Eq.~\ref{eq:thermal_pulse}.

\subsection{Thermistor and biasing circuit models} 

When the thermistor temperature varies, its resistance varies according
to Eq.~\ref{eq:thermistor}:
\begin{equation}
\Delta R(\Delta T) = R_0 \exp\left(\frac{T_0}{T^B + \Delta
T}\right)^\gamma -  R^B
\end{equation}
where $T^B$ is the initial temperature of the bolometer and $R^B = R(T^B)$. 
 Because we do not know the parameters $R_0$, $T_0$ and $T^B$ 
 with the desired accuracy, we adopt an approximation valid for the small $\Delta T$
 that will occur~\cite{Vignati:2010yf}:
\begin{equation}
\label{eq:thermistor_approx} \Delta R(\Delta T) \simeq R^B \left[
\exp(-\eta \Delta T / T^B ) - 1\right]\;,
\end{equation}
where
\begin{equation}
\eta = \left|\frac{d \log{R}}{d \log T} \right|  = {\gamma}\,
\log\frac{R(T)}{R_0}\;.
\end{equation}
$\eta$ is the sensitivity of the thermistor, which has value of order 10 but is not
known with precision.  The advantage of Eq.~\ref{eq:thermistor_approx}
is that the parameter $R^B$ can be measured with precision 
and that the unknown $\eta/T^B$ is just a scale factor applicable to the temperature
variation.  We obtain the expression for the resistance variation after an energy
release by substituting Eq.~\ref{eq:thermal_pulse}
into Eq.~\ref{eq:thermistor_approx} :
\begin{equation}
\Delta R(t)   = R^B\;\left\{ \exp\Big[{-A \left(-e^{-\frac{t}{\tau_r}}
+ \alpha e^{-\frac{t}{\tau_{d1}}} + (1-\alpha) e^{-\frac{t}{\tau_{d2}}}
\right)}\Big]-1\right\} \label{eq:thermistor_pulse}
\end{equation}
where $A = \eta A_T / T^B$.  This expression absorbs the unknown parameter
$\eta/T^B$ with the unknown thermal amplitude
$A_T$.  The thermistor model therefore does not change the number of
unknown parameters and adds the measurable parameter $R^B$.

The relationship between the voltage across the thermistor  $V_{R}(t)$
and its resistance $R(t)$ can be obtained from the differential equation
describing the thermistor's biasing circuit (see Fig.~\ref{fig:biasing}):
\begin{equation}
\bigg[\frac{R_L+R(t)}{R(t)}\bigg]  V_{R}(t)-V_{bias} + R_L c_p
\frac{dV_{R}(t)}{dt} = 0\;. \label{eq:polarization}
\end{equation}
The model we are building is based on variations of the resistance from
the measured value of $R^B$, which in turn generate voltage variations
from  the corresponding voltage $V_R^{B}$:
\begin{equation}
V_R^B = V_{bias} \frac{R^B}{R^B + R_L}\;.
\label{eq:static_voltage}
\end{equation}
By splitting $R(t)$ and $V_R(t)$ into time-independent and time-dependent contributions,
\begin{equation}
R(t) = R^B + \Delta R(t)\quad V_{R}(t) = V^B_{R} + \Delta V_R(t)\;,
\end{equation}
we obtain the differential equation relating resistance and voltage
variations:
\begin{equation}
\Bigg[\frac{R_L+R^B+\Delta R(t)}{R^B+\Delta R(t)}\Bigg] \Bigg[
V_{bias}\frac{R^B}{R^B+R_L} + \Delta V_R(t)\Bigg] -V_{bias} + R_L c_p
\frac{d\Delta V_R(t)}{dt} = 0\;.  \label{eq:voltage_pulse}
\end{equation}
Given the form of $\Delta R(t)$ in Eq.~\ref{eq:thermistor_pulse},
$\Delta V_R(t)$ cannot be obtained in an closed form. In our bolometer model
we solve Eq.~\ref{eq:voltage_pulse} numerically using the Runge-Kutta method~\cite{abramowitz}.

The thermistor and the biasing circuit are the only sources of nonlinearities
of our model, and should be able to describe the nonlinearities observed
in the data.  If we assume that the thermal circuit 
responds linearly, the amplitude $A$ in Eq.~\ref{eq:thermistor_pulse} is directly proportional to
the energy $E$ released in the bolometer,
\begin{equation}
A = c\cdot E\;.
\label{eq:aproptoe}
\end{equation}
The corresponding resistance variation, however, is not proportional to $A$,
because the exponential dependency in Eq.~\ref{eq:thermistor_pulse} does
not transform linearly the shape and the amplitude of the pulse.  Moreover
the voltage variation is not strictly proportional to the resistance variation.
This model description should be
sufficient to generate the shape dependence on energy and the nonlinear
calibration function (Figs.~\ref{fig:shape_std} and \ref{fig:res_std}).
The amplitude dependence on the baseline in Fig.~\ref{fig:stab_std}
can be generated from  Eq.~\ref{eq:thermistor_pulse} by varying the
baseline voltage of the pulse and hence the thermistor resistance $R^B$.

\subsection{Electronics} 

The front-end electronics amplifies the bolometer
signal  $\Delta V_R(t)$ by $G$, a parameter that is measured with
a precision better than $0.2\%$. 
The deviation from linearity of the amplifier, 
in the voltage range we are considering, is less than $0.01\%$.
The output voltage of the amplifier
\begin{equation}
 \Delta V_G(t) = \Delta V_R(t) \cdot G\;,
 \label{eq:amplifier}
 \end{equation}
  is fed into a six-pole Bessel filter, whose transfer function is
\begin{equation}
B(\sigma)=\frac{10395}{\sigma^{6}+21\sigma^{5}+210\sigma^{4}+1260\sigma^{3}+4725\sigma^{2}+10395\sigma+10395}\;.
\label{eq:bessel_h}
\end{equation}
In the above equation $\sigma$ is the normalized Laplace variable which
can be expressed in terms of the frequency $\omega$ as:
\begin{equation}
\sigma = \jmath\omega \frac{2.703395061}{f_b}\,,
\end{equation}
where $f_b$ is the filter cutoff (12\un{Hz} in our case).

The signal is filtered by multiplying its DFT,
$\Delta V_G(\omega)$, by $B(\omega)$, removing the ``DFT wraparound
problem'' with the method described in Ref.~\cite{nr}.  The output of
the filter is then obtained as:
\begin{equation}
\Delta V(t) =  {\cal F}^{-1} [\Delta V_G(\omega)\cdot B(\omega)]\;,
\label{eq:bessel_conv}
\end{equation}
where  ${\cal F}^{-1}$ denotes the inverse DFT.  
\vspace{1cm}

In summary\;, the  model of the signal, from the energy release in the
crystal $E$ to the signal acquired by the ADC $\Delta V(t)$, is obtained
using the thermal model in Eq.~\ref{eq:thermal_pulse}, the thermistor
model in Eq.~\ref{eq:thermistor_pulse}, the voltage across the thermistor
from Eq.~\ref{eq:voltage_pulse}, the amplifier and Bessel filter effects
in Eqns.~\ref{eq:amplifier} and \ref{eq:bessel_conv}:
\begin{equation}
\Delta V(t) = \left\{
\left[E\xrightarrow[\substack{Eq:~\ref{eq:thermal_pulse}\\{\rm analytic}\\{\rm solution}}]{}
\Delta T(t) \xrightarrow[\substack{Eq.~\ref{eq:thermistor_pulse}\\{\rm analytic}\\{\rm solution}}]{}
\Delta R(t) \xrightarrow[\substack{Eq.~\ref{eq:voltage_pulse}\\{\rm numerical\phantom{y}}\\{\rm solution}}]{} 
\Delta V_R(t)\right]
\underset{Eq.~\ref{eq:amplifier}}{\cdot} G
\right\}
\underset{\substack{{Eq.~\ref{eq:bessel_conv}}\\{\rm DFT}\\{\rm convolution}}}{\otimes}
B(t)\;.
\label{eq:signal_model}
\end{equation}

The baseline voltage of the signal, $V^B$, is the sum of two components, the thermistor voltage $V^B_R$ in Eq.~\ref{eq:static_voltage} scaled by the electronics gain $G$, and the offset voltage $V_h$ added by the electronics itself:
\begin{equation}
V^B = V_R^B\cdot G + V_h = V_{bias}G \frac{R^B}{R^B+R_L} + V_h\;.
\label{eq:voltage_baseline}
\end{equation}
To reproduce a real waveform, $V(t)$, we add the baseline voltage to the pulse in Eq.~\ref{eq:signal_model} and we also account for the onset time $t_0$ of the pulse,  which starts about $1\un{s}$ after the beginning of the waveform:
\begin{equation}
V(t) = V^{B} + \Theta(t-t_0)\Delta V(t-t_0)\;,
\label{eq:full_signal_model}
\end{equation}
where $\Theta(t)$ is the Heaviside step function. Since the thermistor temperature is not stable, the resistance $R^B$ and the baseline voltage $V^B$ are not fixed parameters of the model. $V^B$ is measured on data by averaging the first 0.8\un{s} of the waveform. The corresponding value of  $R^B$ is then computed  from Eq.~\ref{eq:voltage_baseline} and used in the signal model.
In Tab.~\ref{tab:modelparams} we list the parameters of the model and indicate
whether they are measured, or determined from fits to signal waveforms.

\begin{table}[htbp]
\centering
\caption{Parameters of the signal model in Eqns.~\protect\ref{eq:signal_model}, \protect\ref{eq:voltage_baseline} and \protect\ref{eq:full_signal_model}.}
\begin{tabular}{clcc}
Parameter & Name & Equation & Estimation \\ 
\hline
$\tau_r$& Thermal rise time & ~\ref{eq:thermal_pulse} & Fit \\
$\alpha$ & Weight of the two thermal decay constants & ~\ref{eq:thermal_pulse} & Fit \\
$\tau_{d1}$ & Fast thermal decay constant& ~\ref{eq:thermal_pulse} & Fit \\
$\tau_{d2}$ & Slow thermal decay constant &~\ref{eq:thermal_pulse} & Fit \\
$c$ & Energy to thermal amplitude conversion  & ~\ref{eq:aproptoe} & Fit \\
$R^B$& Thermistor resistance at the pulse baseline & ~\ref{eq:thermistor_pulse} & Measured \\
$V_{bias}$& Bias voltage & ~\ref{eq:voltage_pulse} & Measured \\
$R_{L}$& Load resistor & ~\ref{eq:voltage_pulse} & Measured \\
$c_{p}$& Parasitic capacitance & ~\ref{eq:voltage_pulse} & Measured \\
$G$ & Electronics  gain&~\ref{eq:amplifier} & Measured \\
$f_b$ & Bessel filter cutoff frequency &~\ref{eq:bessel_conv} & Measured \\
$V_h$ & Electronics  offset voltage&~\ref{eq:voltage_baseline} & Measured \\
$V^B$ & Baseline voltage &~\ref{eq:full_signal_model} & Measured \\
$t_0$ & Onset time of the  pulse &~\ref{eq:full_signal_model} & Fit \\
\hline
\end{tabular}
\label{tab:modelparams}
\end{table}

\section{Estimation of the signal model}\label{sec:pulsefit}

We intend that the model we developed  accounts for all the
nonlinearities of the signal. The unknown thermal parameters in Tab.~\ref{tab:modelparams} 
are expected to be independent of the energy, except for the amplitude $A$, which
should be proportional to the energy (Eq.~\ref{eq:aproptoe}).
The energy independence allows us to determine unmeasured parameters
from fits to pulses at a single energy, and apply the resulting model over the entire
range of energies.
We performed fits on  particle pulses occurring in an energy
window of 30\un{keV} around the 2615\un{keV} $\gamma$ peak. We performed a separate
set of fits on heater pulses, because their shape
differs from the shape of particle pulses.

\begin{figure}[tbp]
\begin{center}
\includegraphics[width=0.9\textwidth,clip=false]{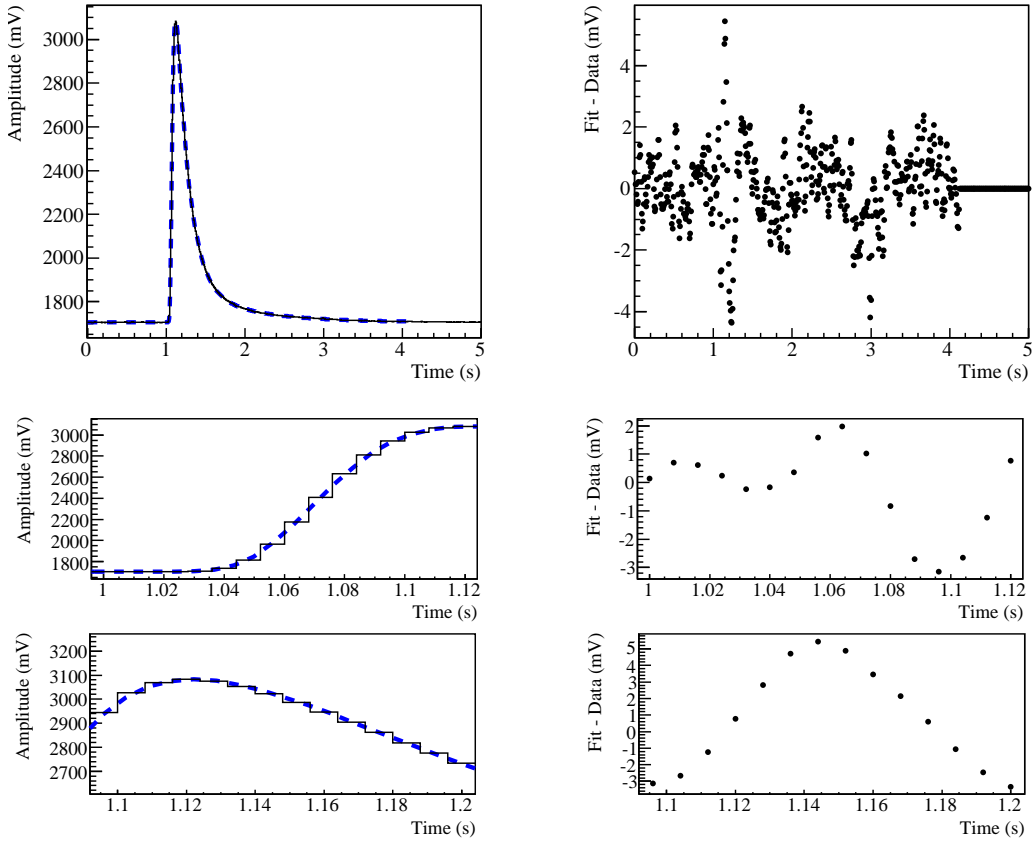}
\caption{Fit of a 2615\un{keV} pulse. Data (black solid lines) with superimposed fit function (Eq.~\protect\ref{eq:full_signal_model})
(blue dashed lines) are shown in the left column, fit residuals are shown in the right column. The two bottom
rows display a zoom of the rise and of the maximum of the pulse, respectively. The fit region ends
when the pulse falls to within 3 standard deviations of the baseline level.}
\label{fig:pulsefit}
\end{center}
\end{figure}

Figure~\ref{fig:pulsefit} shows a typical fit to a 2615\un{keV}$\gamma$-ray pulse, Tab.~\ref{tab:pulsefit} reports the parameters averaged over 25 fits of particle and heater pulses.
\begin{table}[tbp]
\centering
\caption{Average parameters fitted on particle (2615\un{keV} $\gamma$) and heater pulses and the $\chi^2/{\rm ndf}$ of the fits.} 
\begin{tabular}{lcc}
Parameter &  Particle & Heater \\ 
\hline
$\tau_{r}\;\un{(ms)}$    & $20.70 \pm 0.05$         & $18.8 \pm 0.1$\\
$\alpha\;$               & $0.916 \pm 0.002$        & $0.911 \pm 0.002$\\
$\tau_{d1}\;\un{(ms)}$   & $158.1 \pm 0.5$          & $187.3 \pm 0.6$\\
$\tau_{d2}\;\un{(ms)}$   & $770 \pm 20$             & $970 \pm 20$\\
$c\;\un{(1/MeV)}$        & $0.04703 \pm 0.00006$    & $0.04356 \pm 0.00009$\\
$t_0\;\un{(s)}$          & $1.0145 \pm 0.0005$      & $0.9944 \pm 0.0005$\\
$\chi^2/{\rm ndf}\;$     & $2.4 \pm 0.3$            & $2.0 \pm 0.2$\\
\hline
\end{tabular}
\label{tab:pulsefit}
\end{table}
The average $\chi^2$ is about twice the number of degrees of freedom, i.e. about
twice the value one expects for a perfect model.
Shortcomings of the model are also apparent in the fit residuals on the right of the figure,
where  mismatches are evident in the leading edge  and in the vicinity of the maximum of the pulse. 
The voltage amplitude of the fit function is slightly biased,
and found to be, on average, higher than the amplitude of the pulse by
$0.17 \pm 0.02\un{\%}$ for particle pulses and  $0.37 \pm 0.02\un{\%}$ for particle heater pulses.

The fit error can be ascribed to an incompleteness of the thermal model,
that would probably benefit from the inclusion of second order effects like the nonlinearities
of the thermal elements and the electrothermal feedback. 
For our purposes, however, the model performs  well, reproducing the signal
at the per mil level. 

\section{Noise generation}
A complete and predictive model for the noise of CUORE bolometers is missing. The main sources of fluctuations
that spoil the energy resolution have different origins and,  to estimate 
the overall contribution, each of them should be 
propagated with the transfer function of each step of the acquisition chain. 
Examples of these sources are the Johnson noise of the load resistors,
 vibrations of the experimental apparatus that dissipate  energy in the bolometer and 
instabilities of the cryostat temperature. All these effects contribute to the power spectrum shown in Fig.~\ref{fig:noise_std}.

Although the noise from the load resistors and amplifiers is predictable, a model
for the noise from vibrations is not in hand.
Moreover, it varies from bolometer to bolometer, because of the weak reproducibility of the assembly.
On this account we adopted  a  
statistical model for the noise. We simply require that, on average, the simulated time series
behave like the experimental one, namely that the average power spectrum of the simulated baselines is as close as possible to data.

We followed an approach \cite{Carrettoni20101982} based on an application of the Carson's theorem \cite{Carson} to a discrete time series. 
A random waveform $n(t_i)$ can be represented as a superposition of independent pulses of fixed shape $g(t_i)$ and amplitude $A$, 
distributed in time according to a Poisson process of rate $\lambda$:
\begin{equation}
n(t_i)=A\,\sum_l g(t_i-t_l)\;,
\label{eq:pulse_train}
\end{equation}
where the differences between consecutive $t_l$'s follow an exponential distribution with mean $1/\lambda$. The theorem states that  $N(\omega_k)$ and $G(\omega_k)$, the average 
power spectra of $n(t_i)$ and $g(t_i)$, respectively  (see Eq.~\ref{eq:nps}), satisfy the relationship: 
\begin{equation} 
N(\omega_k)  =  \lambda T A^2 G(\omega_k)\;,
\label{eq:carson}
\end{equation}
where $T$ is the length of the time series (5.008\un{s} in our case).
To utilize the theorem one must find a shape $g(t_i)$ for which the power spectrum is proportional to the average
power spectrum of the noise time series to be simulated. $A^2$ and $\lambda$ are 
adjustable  parameters with product fixed by Eq.~\ref{eq:carson}.  These parameters set the aspect of the noise in the time domain,
i.e. the same power spectrum can be produced with a small rate of pulses with large amplitude
and vice versa.  Once $g(t_i)$ and, say, $\lambda$ are determined the method is fully specified
 and $n(t_i)$ is  obtained from Eq.~\ref{eq:pulse_train}.
 
We built $g(t_i)$ as the inverse DFT of
\begin{equation} 
 g(\omega_k) = \sqrt{N(\omega_k)}e^{\jmath\theta_k}\;,
\end{equation} 
where $\theta_k$ is a phase randomly sampled within $[0,2\pi]$. The reality of $g(t_i)$ is guaranteed by imposing the constraint $g(\omega_k) = g^*(-\omega_k)$. $\lambda$
was chosen equal to the Nyquist frequency (62.5\un{Hz}), i.e.  the maximum rate producing
distinguishable pulses.

\begin{figure}[htbp]
\centering
\begin{minipage}{0.48\textwidth}
\includegraphics[clip=true,width=1\textwidth]{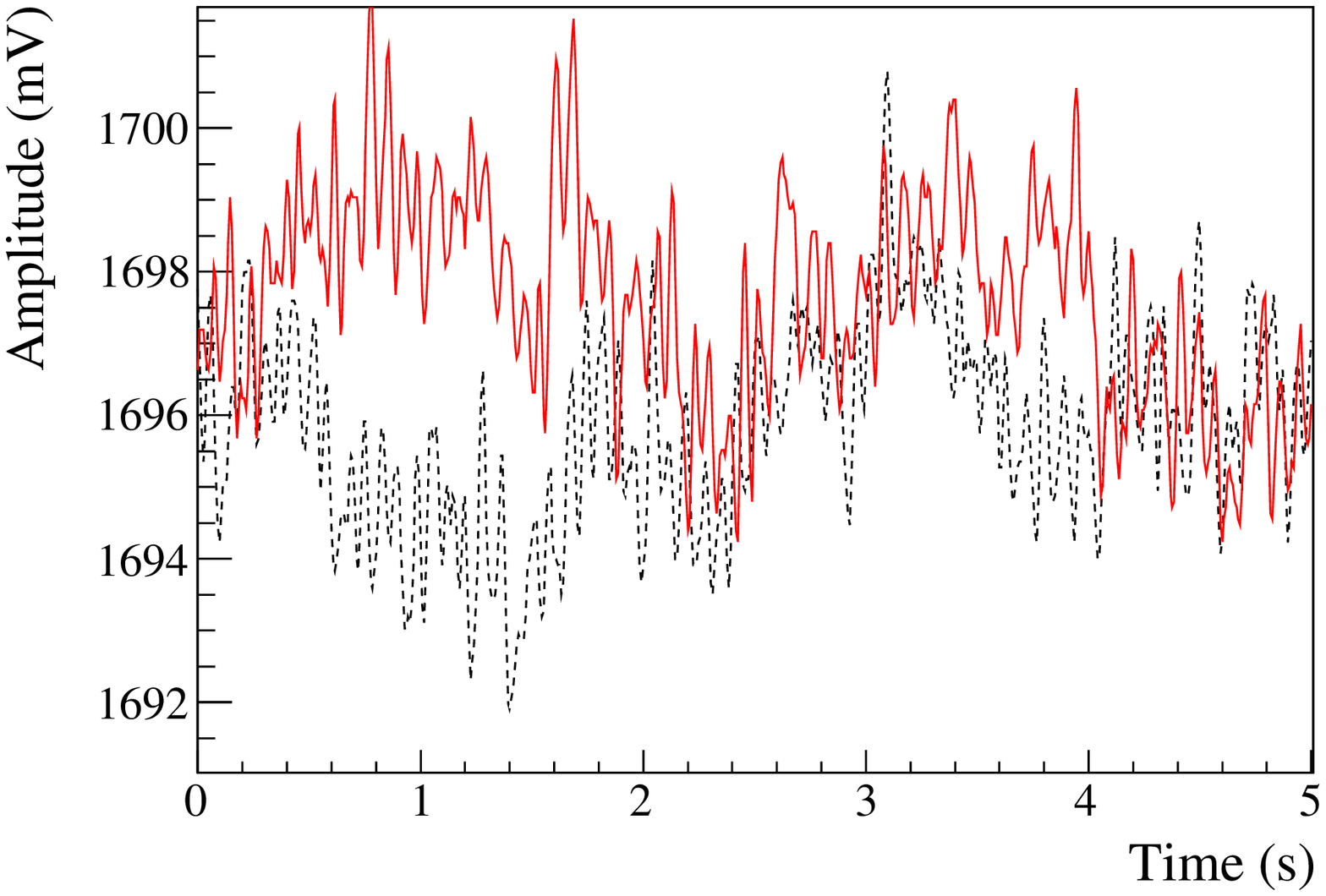}  
\end{minipage}\hfill
\begin{minipage}{0.48\textwidth}
\includegraphics[clip=true,width=1\textwidth]{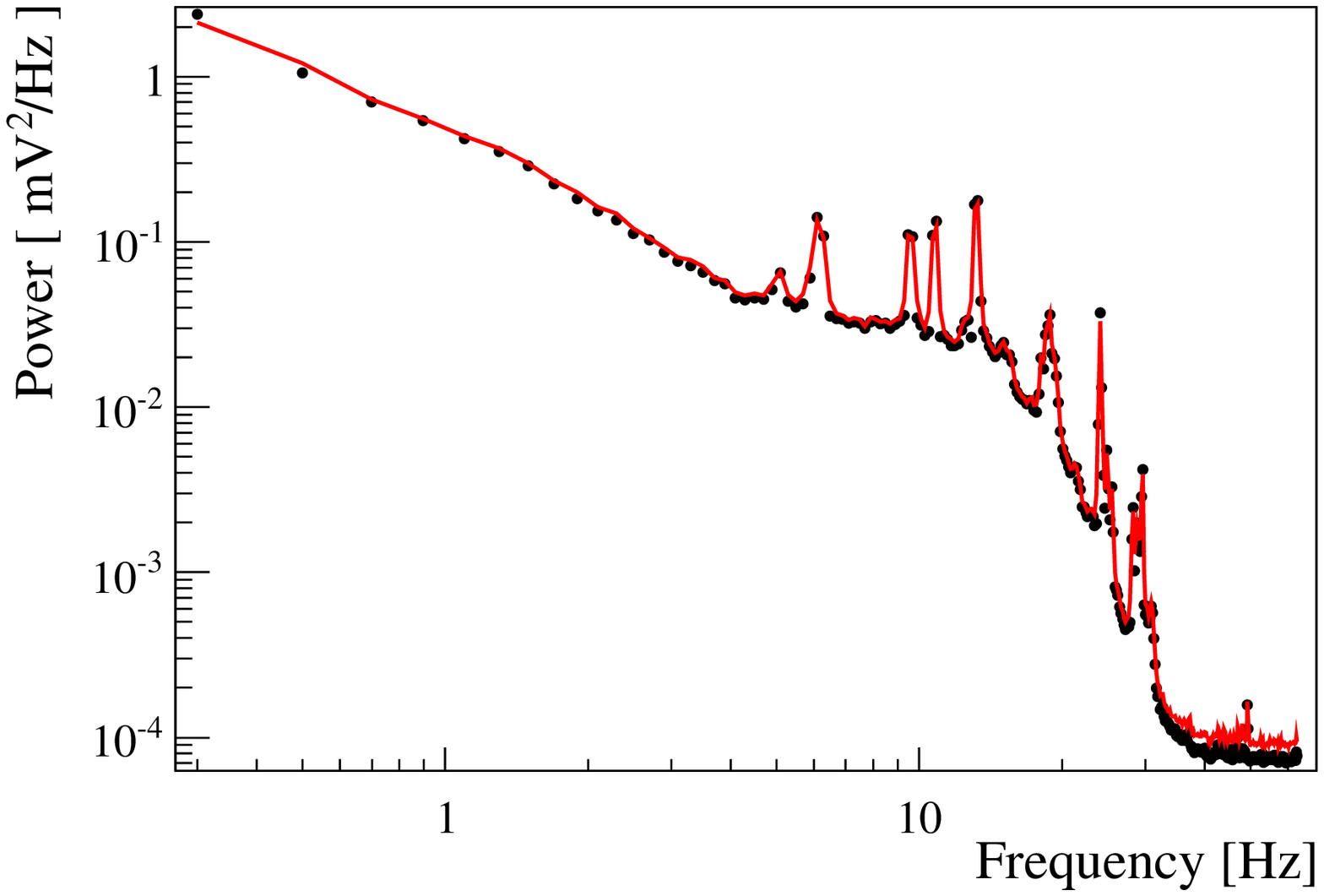}  
\end{minipage}
\caption{Comparison of noise simulation (solid red line) with data (dashed black line / dots). Sample waveforms (left) and
the power spectrum estimated from a large number of waveforms (right).}
\label{fig:noise_std_mc}
\end{figure}

The left panel of Fig.~\ref{fig:noise_std_mc} shows a simulated and a measured noise waveform, and their appeareance is similar. The right panel of the figure
shows the agreement between simulated and measured power spectra. 

\section{Simulation and validation with data} 

We built a simulation engine that is able to generate particle, heater,
and pure noise waveforms. It uses as input the parameters of the signal
model in Tab.~\ref{tab:modelparams} and the measured noise power spectrum. 
To correct the small error of the model in reproducing the pulse shape (see Sec.~\ref{sec:pulsefit}), the engine scales the simulated signals by the estimated bias, so as
to match the amplitude of measured signals. The simulated noise is summed to
the signal assuming that it is purely additive. 
The additivity is supported by the fact that the absolute
energy resolution for heater pulses is independent of the
energy release, and by the residuals in Fig.~\ref{fig:pulsefit} that
do not seem correlated with the time evolution of the pulse.

The simulation can be highly customized. 
One can choose the energy distribution of the events to be generated, 
the baseline distribution, and the distribution of the time
interval between events. The features
of the signal described in Sec.~\ref{sec:snfeatures} should be automatically
reproduced by the model, and signals close in time
are summed to reproduce pileups.

To validate the simulation engine, we show the results
of a simulation configured to reproduce the data shown in
Sec.~\ref{sec:snfeatures}. Signals were generated sampling their
energy from the spectrum in Fig.~\ref{fig:energy_spectrum}, the
baseline was generated in the range of Fig.~\ref{fig:stab_std}, the
time delay between particle pulses was generated following an
exponential distribution with mean $1/(133\un{mHz})$, and heater pulses
were generated every 300\un{s}. Particle and heater pulses were
generated using the fitted parameters in Tab.~\ref{tab:pulsefit}, and the noise was generated from the power spectrum in Fig.~\ref{fig:noise_std}.

\begin{figure}[hbp]
\centering
\begin{minipage}{0.48\textwidth}
\includegraphics[clip=true,width=1\textwidth]{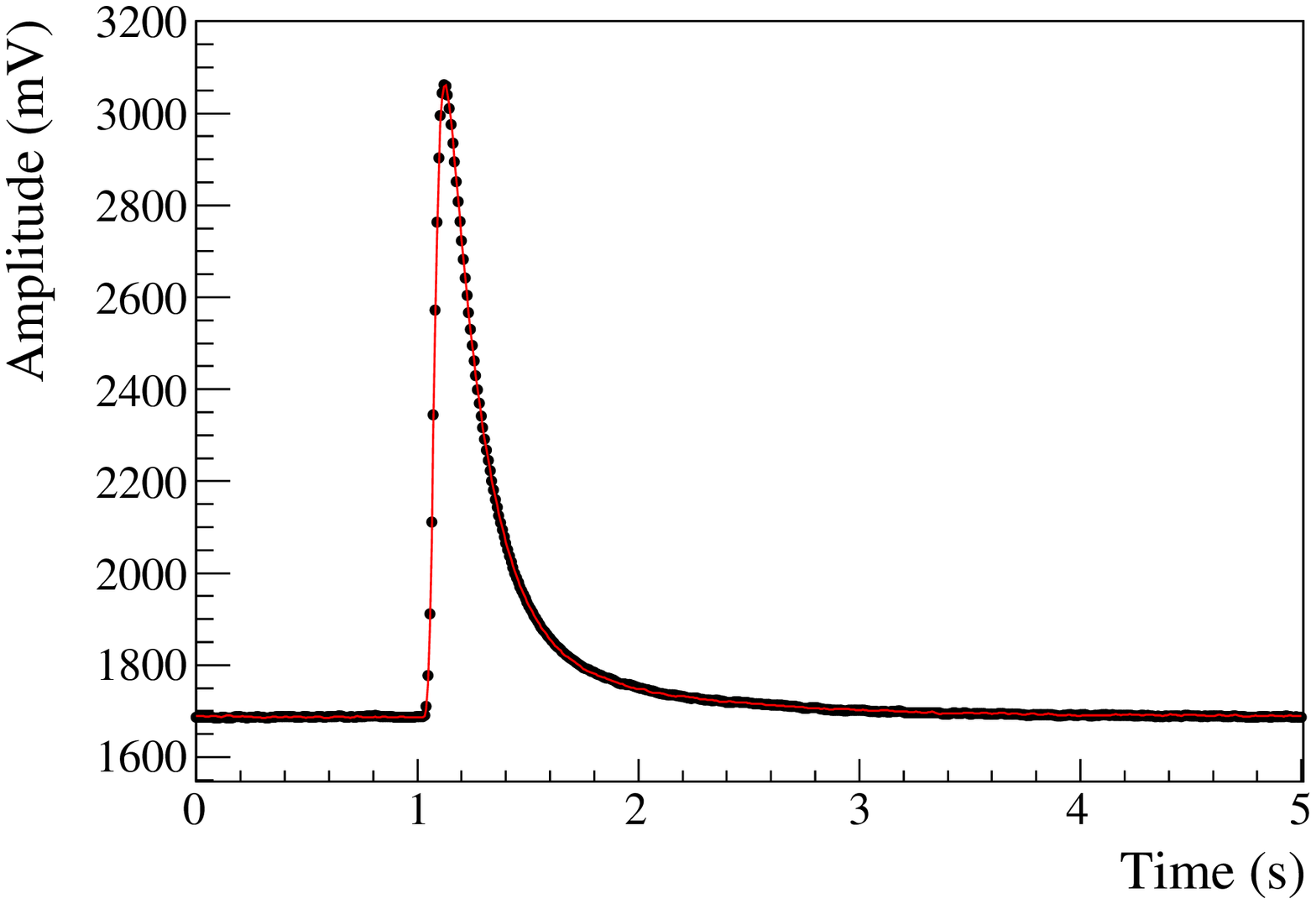}
\end{minipage}\hfill
\begin{minipage}{0.48\textwidth}
\includegraphics[clip=true,width=1\textwidth]{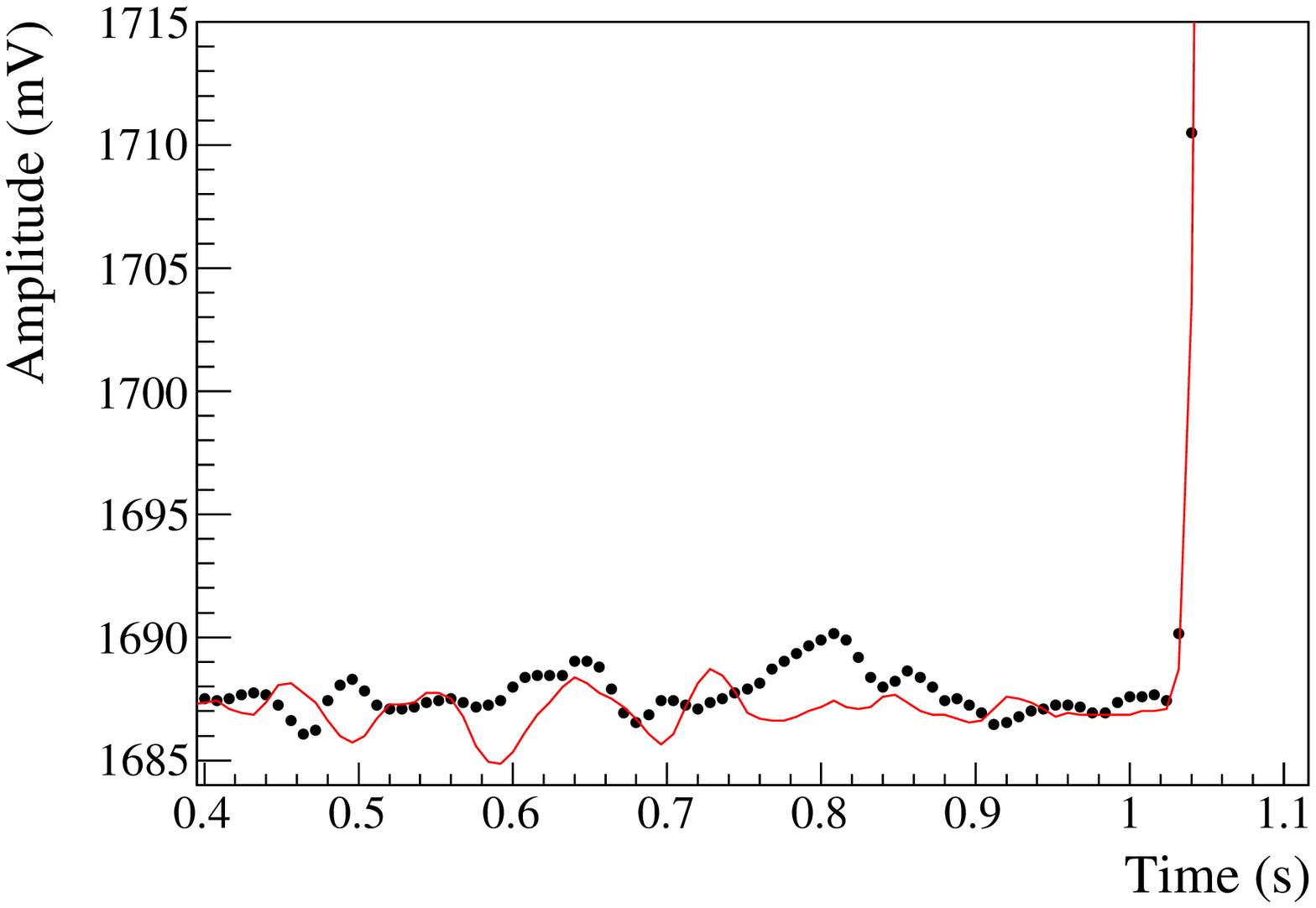}
\end{minipage}
\caption{Comparison of the simulation (solid red line) of a 2615\un{keV} $\gamma$ pulse
with data (black dots). Full waveform (left) and expansion of the leading edge of the signal (right).}
\label{fig:pulses_std_mc}
\end{figure}

\begin{figure}[hbp]
\centering
\begin{minipage}{0.48\textwidth}
    \begin{overpic}[clip=true,width=1\textwidth]{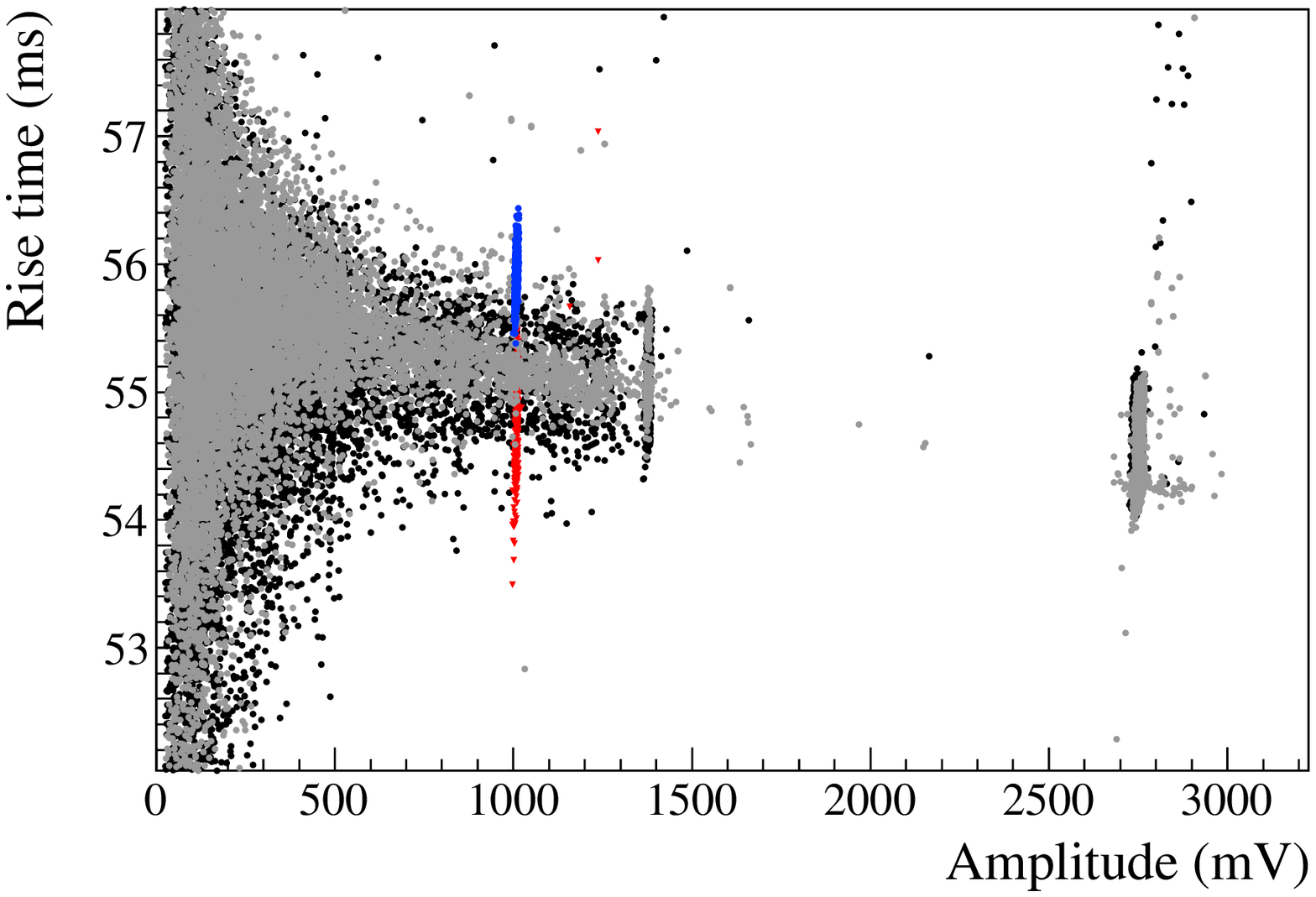}  
      \put(34,20){\footnotesize heater}
      \end{overpic}
\end{minipage}\hfill
\begin{minipage}{0.48\textwidth}
    \begin{overpic}[clip=true,width=1\textwidth]{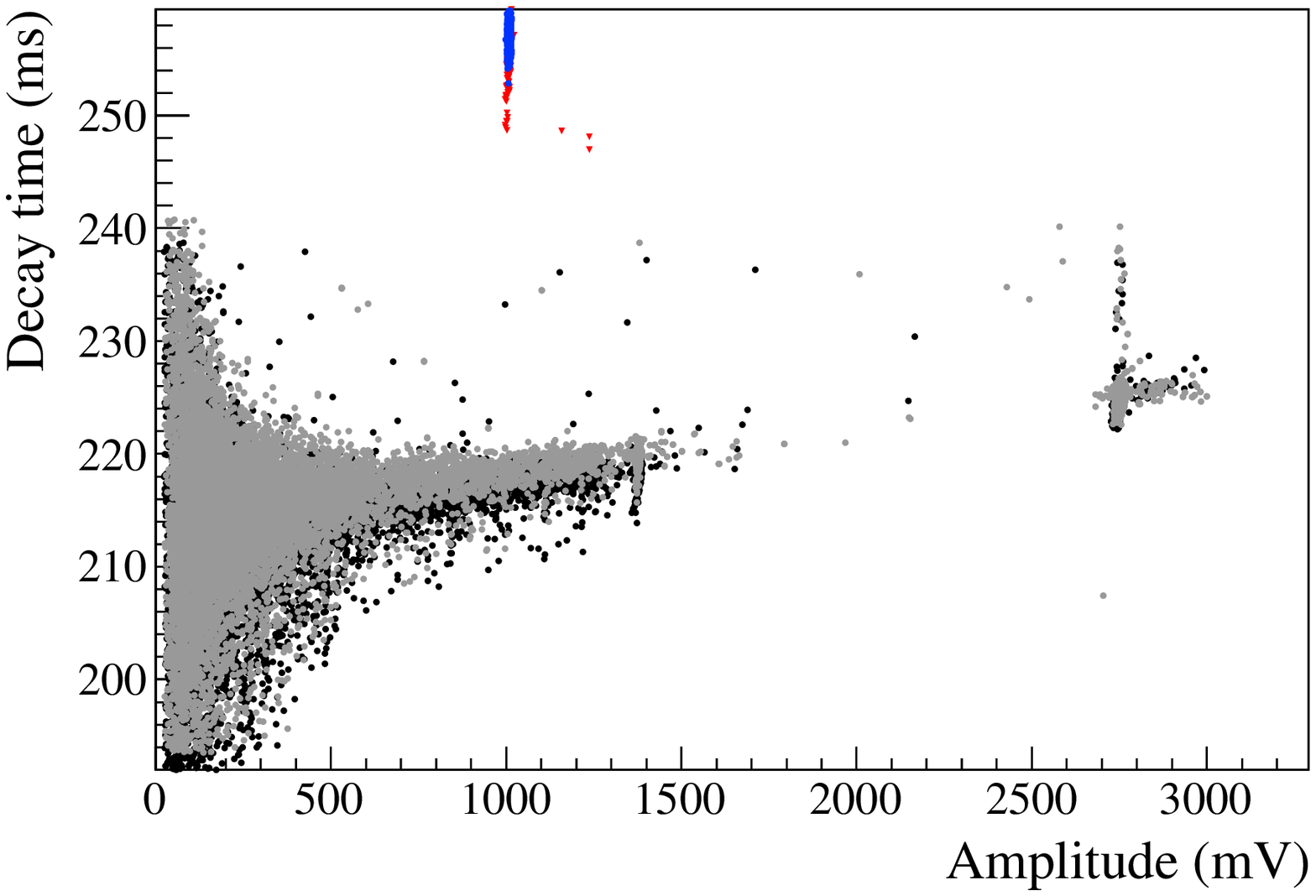}  
        \put(42,60){\footnotesize heater}
    \end{overpic}
\end{minipage}
\caption{Comparison of the rise time (left) and decay time (right). The
black, gray, red, and blue dots represent measured particle pulses,
simulated particle pulses,
measured heater pulses,
and simulated heater pulses respectively.}
 \label{fig:shape_std_mc}
\end{figure}

\begin{figure}[htbp]
\begin{center}
\includegraphics[clip=true,width=0.5\textwidth]{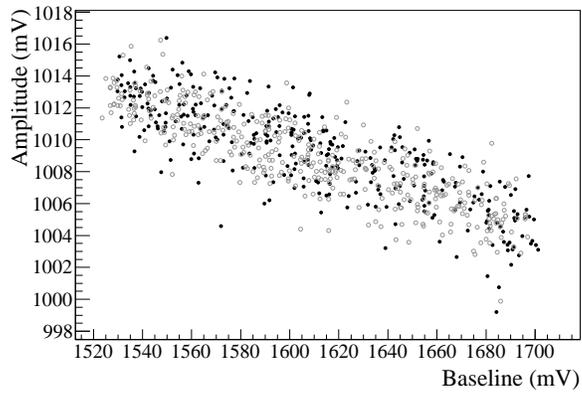}
\caption{Comparison of the amplitude of heater pulses versus baseline from data (solid black)
and simulation (open gray).} \label{fig:stab_std_mc}
\end{center}
\end{figure}

\begin{figure}[tbp]
\begin{center}
\begin{minipage}{0.48\textwidth}
\includegraphics[clip=true,width=1\textwidth]{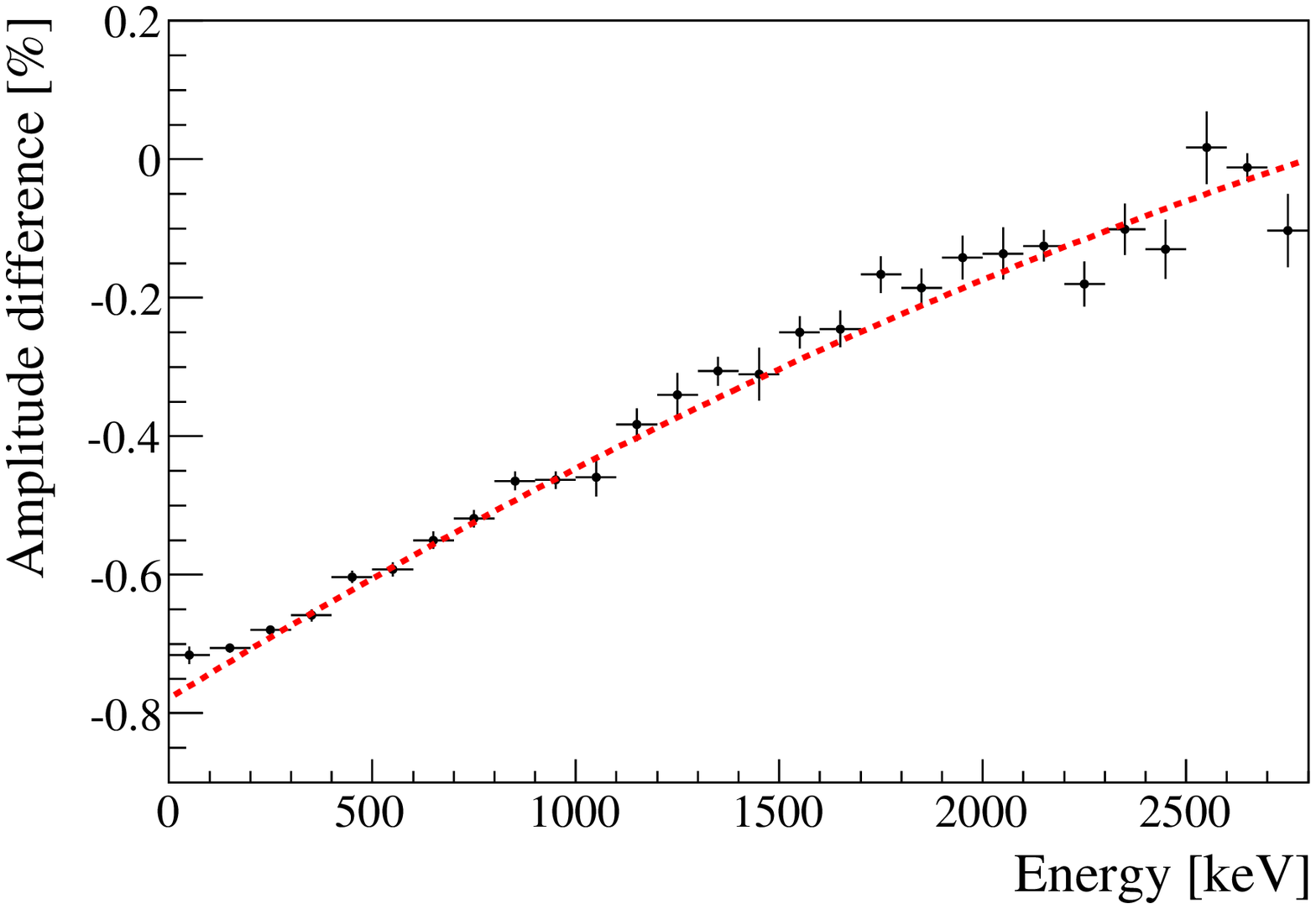}
\end{minipage}\hfill
\begin{minipage}{0.48\textwidth}
\includegraphics[clip=true,width=1\textwidth]{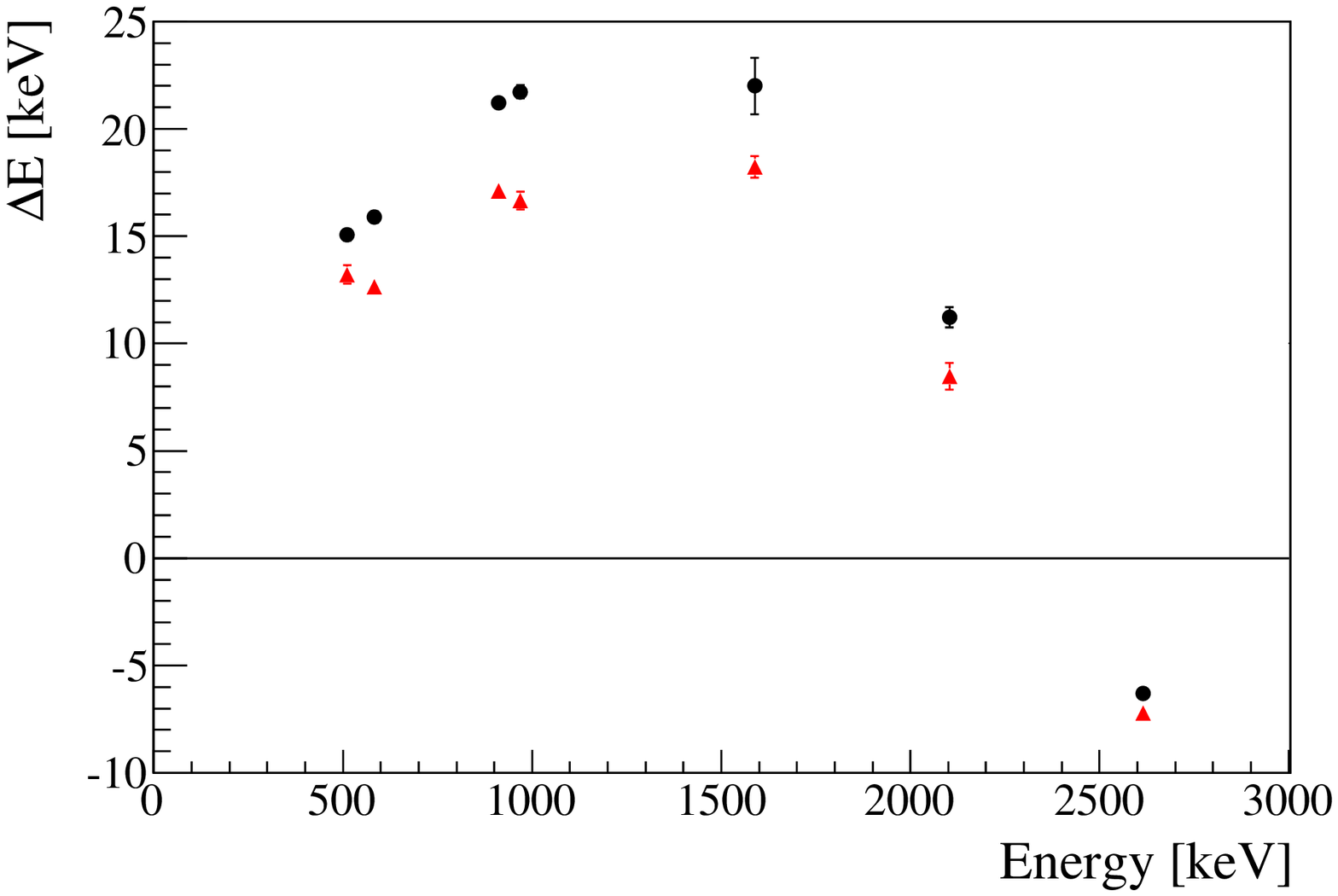}
\end{minipage}
\caption{
Difference between                     
the amplitudes of simulated and real pulses as a function of the energy (left), and 
deviation from linearity of the calibration function (right) for data (black circles) and simulation (red triangles).}
\label{fig:res_std_mc}
\end{center}
\end{figure}

The comparison of the simulation with the data shows good agreement.
The shape of the pulses with  noise added reproduces well the acquired 
waveforms (Fig.~\ref{fig:pulses_std_mc}). 
The distributions of the rise and decay times
shown in Fig.~\ref{fig:shape_std_mc} also confirm the agreement. We attribute a small mismatch 
in the rise time of heater pulses to imperfections of the model (see
Sec.~\ref{sec:pulsefit}).  The correlation between pulse amplitude and        
baseline is very well reproduced (Fig.~\ref{fig:stab_std_mc}). 
The dependence of amplitude on energy follows  the data well at high energies (Fig.~\ref{fig:res_std_mc}).
The mismatch of order 1\% at zero energy is another manifestation of the imperfections
of the model. 
In most applications this error is ignorable, nonetheless we incorporated an option
to generate waveforms by sampling from the amplitude spectrum instead
of the energy spectrum, in which case the mismatch is suppressed by fiat.%

\newpage
\cleardoublepage
\section{Applications}
The simulation can be used to test and tune analysis algorithms, comparing the results
with the so called ``Monte Carlo truth'', i.e. the generated values of  baseline, amplitude, energy
and the time of each signal.  In same cases, in fact, the signal identification is not straightforward
and the energy could be wrongly estimated.

\begin{figure}[bp]
\begin{center}
\includegraphics[clip=true,width=0.5\textwidth]{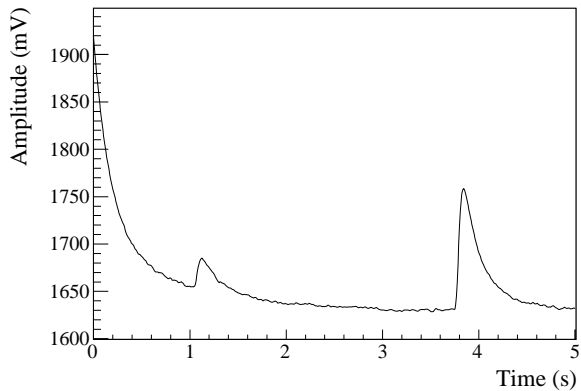}
\caption{Simulated pileups. The pulse at $\sim 1\un{s}$ was generated with an energy of 63\un{keV},
lies on the tail of a previous pulse of 1899\un{keV} and is succeeded by a pulse of 263\un{keV}.} 
\label{fig:pileups_mc}
\end{center}
\end{figure}

For example the data analysis can be complicated by pileups, that alter the baseline and the shape of the signals.
A simulated sequence of close signals is shown in Fig.~\ref{fig:pileups_mc},
where it can be seen how a signal can be modified by other signals, and how its identification
is complicated. In this case the simulation can be used to improve the analysis algorithms and to estimate
the error on the results. 

\begin{figure}[htbp]
\centering
\begin{minipage}{0.48\textwidth}
\includegraphics[clip=true,width=1\textwidth]{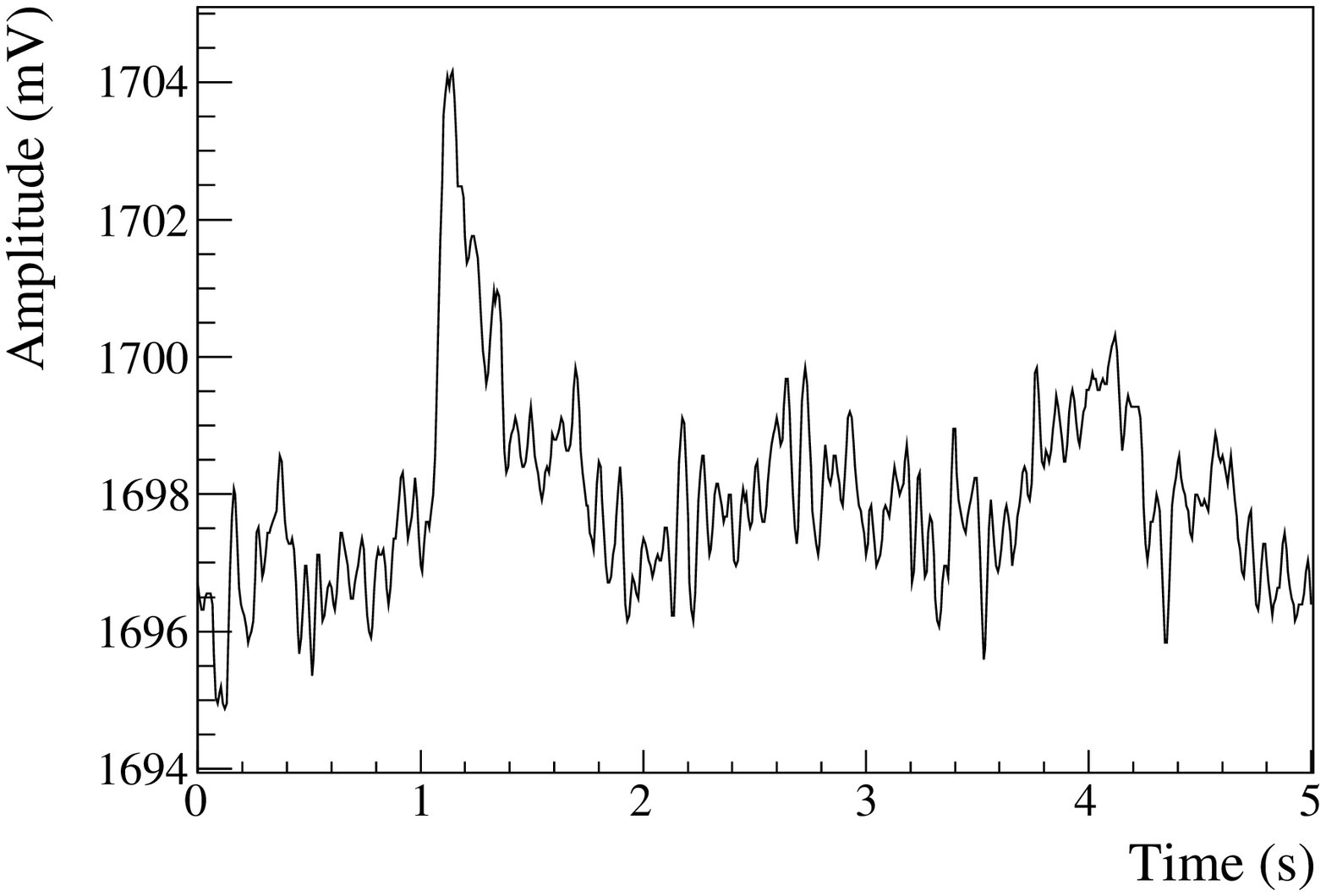}  
\end{minipage}
\hfill
\begin{minipage}{0.48\textwidth}
\includegraphics[clip=true,width=1\textwidth]{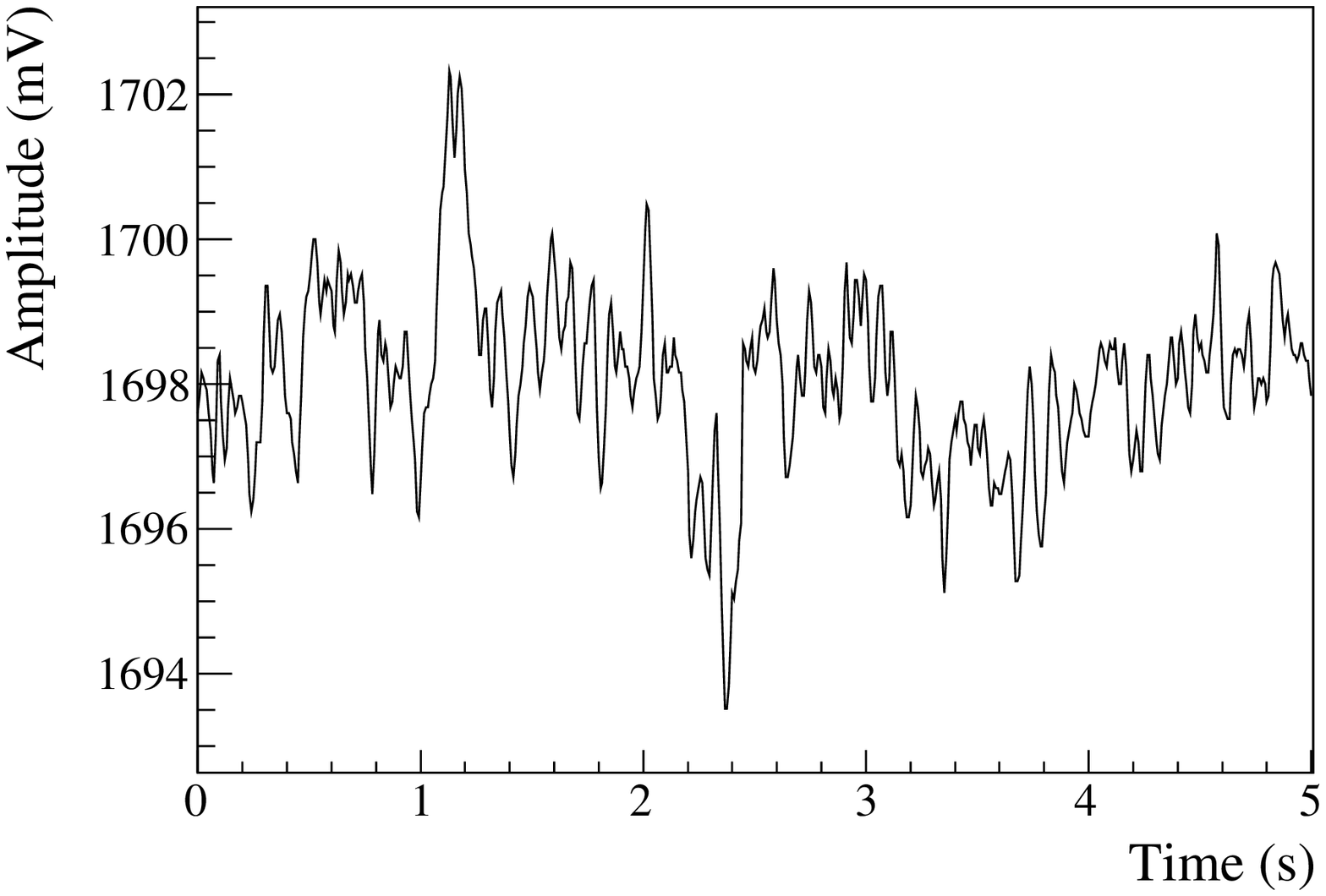}  
\end{minipage}
\caption{12\un{keV} (left) and 6\un{keV} (right) simulated pulses.} 
\label{fig:lowenergy_mc}
\end{figure}

A potential application is the estimation of  detection efficiencies at low energy.
Recently it has been demonstrated~\cite{DiDomizio:2010ph} that CUORE may
have an energy threshold of only a few keV, thus being sensitive to Dark Matter interactions
and rare nuclear decays. Figure~\ref{fig:lowenergy_mc} shows simulations
of very small pulses. In this regime noise can both mask signal and mimic signal.
Simulations produced by our engine may be used to test the immunity
of analysis algorithms to both types of error. 
When a heater is available it is used to produce controlled pulses and estimate the
detection efficiencies. The simulation can correct for the difference in pulse shape
between particle and heater generated energy deposit. With no heater the simulation
could serve on its own to estimate the detection efficiencies.

\acknowledgments
We thank the members of the CUORE collaboration, in particular G.~Pessina 
for fruitful discussions on the signal model and on the electronics setup,
and O.~Cremonesi for suggestions on the noise generation. We thank F.~Bellini, P.~Decowski, F.~Ferroni, F.~Orio, C.~Rosenfeld and C.~Tomei  for comments
on the manuscript.
\bibliographystyle{JHEP} 
\bibliography{main}

\providecommand{\href}[2]{#2}\begingroup\raggedright\begin{thebibliography}{10}

\bibitem{Ardito:2005ar}
R.~Ardito {\em et.~al.}, {\it {CUORE: A cryogenic underground observatory for
  rare events}},  {\em arXiv:hep-ex/0501010} (2005)
  [\href{http://xxx.lanl.gov/abs/hep-ex/0501010}{{\tt hep-ex/0501010}}].

\bibitem{ACryo}
C.~Arnaboldi {\em et.~al.}, {\it {CUORE}: {A} {C}ryogenic {U}nderground
  {O}bservatory for {R}are {E}vents},  {\em {N}ucl. {I}nstr. {M}eth. in {P}hys.
  {R}es. A.} {\bf 518} (2004) 775,
  [\href{http://xxx.lanl.gov/abs/hep-ex/0212053v1}{{\tt hep-ex/0212053v1}}].

\bibitem{DiDomizio:2010ph}
S.~Di~Domizio, F.~Orio, and M.~Vignati, {\it {Lowering the energy threshold of
  large-mass bolometric detectors}},  {\em JINST} {\bf 6} (2011) P02007,
  [\href{http://xxx.lanl.gov/abs/1012.1263}{{\tt arXiv:1012.1263}}].

\bibitem{Vignati:2010yf}
M.~Vignati, {\it {Model of the Response Function of Large Mass Bolometric
  Detectors}},  {\em J.Appl.Phys.} {\bf 108} (2010) 084903,
  [\href{http://xxx.lanl.gov/abs/1006.4043}{{\tt arXiv:1006.4043}}].

\bibitem{Bellini:2010iw}
F.~Bellini {\em et.~al.}, {\it {Response of a $TeO_{2}$ bolometer to alpha
  particles}},  {\em JINST} {\bf 5} (2010) P12005,
  [\href{http://xxx.lanl.gov/abs/1010.2618}{{\tt arXiv:1010.2618}}].

\bibitem{wang}
N.~Wang, F.~C. Wellstood, B.~Sadoulet, E.~E. Haller, and J.~Beeman, {\it
  Electrical and thermal properties of neutron-transmutation-doped {G}e at 20
  m{K}},  {\em Phys. Rev. B} {\bf 41} (1990), no.~6 3761.

\bibitem{Itoh}
K.~M. {Itoh} {\em et.~al.}, {\it Neutron transmutation doping of isotopically
  engineered {G}e},  {\em {A}ppl. {P}hys. {L}ett.} {\bf 64} (1994) 2121.

\bibitem{stabilization}
A.~Alessandrello {\em et.~al.}, {\it Methods for response stabilization in
  bolometers for rare decays},  {\em {N}ucl. {I}nstr. {M}eth. in {P}hys. {R}es.
  A.} (1998), no.~412 454.

\bibitem{Arnaboldi:2003yp}
C.~Arnaboldi, G.~Pessina, and E.~Previtali, {\it {A programmable calibrating
  pulse generator with multi- outputs and very high stability}},  {\em IEEE
  Trans. Nucl. Sci.} {\bf 50} (2003) 979.

\bibitem{Mott:1969}
N.~F. Mott, {\it Localized states in a pseudogap and near extremities of
  conduction and valence bands},  {\em {P}hil. {M}ag.} {\bf 19} (1969) 835.

\bibitem{efros}
A.~Efros and B.~Shklovskii, {\em Electronic properties of doped
  semiconductors}, p.~202.
\newblock Springer-Verlag, Berlin, 1984.

\bibitem{Itoh:1996}
K.~M. Itoh {\em et.~al.}, {\it Hopping conduction and metal-insulator
  transition in isotopically enriched neutron-transmutation-doped
  $^{70}${G}e:{G}a},  {\em Phys. Rev. Lett.} {\bf 77} (1996), no.~19 4058.

\bibitem{AProgFE}
C.~Arnaboldi {\em et.~al.}, {\it The programmable front-end system for
  {CUORICINO}, an array of large-mass bolometers},  {\em IEEE Trans. Nucl.
  Sci.} {\bf 49} (2002) 2440.

\bibitem{ioanprod}
C.~Arnaboldi {\em et.~al.}, {\it Production of high purity {T}e{O}$_2$ single
  crystals for the study of neutrinoless double beta decay},  {\em J. Cryst.
  Growth} {\bf 312} (2010), no.~20 2999.

\bibitem{pessinamod}
A.~Alessandrello {\em et.~al.}, {\it {A}n {E}lectrothermal {M}odel for {L}arge
  {M}ass {B}olometric {D}etectors},  {\em IEEE {T}rans. {N}ucl. {S}ci.} {\bf
  40} (1993), no.~4 649.

\bibitem{probst:1995}
F.~Pr\"{o}bst {\em et.~al.}, {\it Model for cryogenic particle detectors with
  superconducting phase transition thermometers},  {\em J. Low. Temp. Phys.}
  {\bf 100} (1995) 69--104.

\bibitem{abramowitz}
M.~Abramowitz and I.~A. Stegun, {\em Handbook of Mathematical Functions with
  Formulas, Graphs, and Mathematical Tables}.
\newblock Dover, New York, ninth dover printing, tenth gpo printing~ed., 1964.

\bibitem{nr}
W.~H. Press, S.~A. Teukolsky, W.~T. Vetterling, and B.~P. Flannery, {\em
  Numerical recipes in C (2nd ed.): the art of scientific computing}.
\newblock Cambridge University Press, New York, NY, USA, 1992.

\bibitem{Carrettoni20101982}
M.~Carrettoni and O.~Cremonesi, {\it Generation of noise time series with
  arbitrary power spectrum},  {\em Comput. Phys. Commun.} {\bf 181} (2010),
  no.~12 1982.

\bibitem{Carson}
J.~Carson, {\it {The Statistical Energy-Frequency Spectrum of Random
  Disturbances}},  {\em Bell Syst. Techn. J.} {\bf 10} (1931) 374.

\end{thebibliography}\endgroup

%
\end{document}